\documentclass[letterpaper,twocolumn,10pt]{article}
\usepackage{usenix2019_v3}
\microtypecontext{spacing=nonfrench}

\usepackage{balance}

\usepackage{enumitem}

\usepackage{amsmath,amsfonts}
\usepackage{algorithmic}
\usepackage{graphicx}
\usepackage{textcomp}
\usepackage{bm}
\usepackage{xspace}
\usepackage{cleveref}
\usepackage{booktabs}
\usepackage{multirow}

\usepackage[printwatermark]{xwatermark}

\usepackage{hyperref}

\AtBeginDocument{%
	\Crefformat{section}{\S#2#1#3}%
	\crefformat{section}{\S#2#1#3}%
	\Crefformat{subsection}{\S#2#1#3}%
	\crefformat{subsection}{\S#2#1#3}%
	\Crefformat{subsubsection}{\S#2#1#3}%
	\crefformat{subsubsection}{\S#2#1#3}%
	\Crefrangeformat{section}{\S\S#3#1#4 to~#5#2#6}%
	\crefrangeformat{section}{\S\S#3#1#4 to~#5#2#6}%
	\Crefmultiformat{section}{\S\S#2#1#3}{ and~#2#1#3}{, #2#1#3}{ and~#2#1#3}%
	\crefmultiformat{section}{\S\S#2#1#3}{ and~#2#1#3}{, #2#1#3}{ and~#2#1#3}%
}

\usepackage[numbers,sort]{natbib}
\usepackage{caption}
\usepackage{subcaption}

\usepackage{xargs}                      

\usepackage[colorinlistoftodos,prependcaption,textsize=tiny,textwidth=15mm]{todonotes}
\newcommandx{\unsure}[2][1=]{\todo[linecolor=red,backgroundcolor=red!25,bordercolor=red,#1]{#2}}
\newcommandx{\change}[2][1=]{\todo[linecolor=blue,backgroundcolor=blue!25,bordercolor=blue,#1]{#2}}
\newcommandx{\info}[2][1=]{\todo[linecolor=OliveGreen,backgroundcolor=OliveGreen!25,bordercolor=OliveGreen,#1]{#2}}
\newcommandx{\improve}[2][1=]{\todo[linecolor=Plum,backgroundcolor=Plum!25,bordercolor=Plum,#1]{#2}}

\graphicspath{ {./images/} }

\newcommand{\eg}{e.g.,\xspace} %
\newcommand{\ie}{i.e.,\xspace} %
\newcommand{\vs}{vs.\xspace} %

\newcommand{\seqT}{\text{\bf T}}%

\newcommand{\tool}{\textsc{GAME-UP}\xspace}%
\newcommand{\drebin}{DREBIN\xspace}%
\newcommand{\drebinnew}{DREBIN20\xspace}%
\newcommand{\sleipnir}{SLEIPNIR\xspace}%
\newcommand{\ember}{EMBER\xspace}%
\newcommand{\androzoo}{AndroZoo\xspace}%
\newcommand{\tesseract}{\textsc{Tesseract}\xspace}%

\renewcommand{\paragraph}[1]{{\vskip 0pt \noindent\textbf{#1.} }}

\begin{document}

\title{Realizable Universal Adversarial Perturbations for Malware}

\author{
  {\rm 
    Raphael~Labaca-Castro\footnotemark[2]~,~%
    Luis~Muñoz-González\footnotemark[5]~,~%
    Feargus~Pendlebury\footnotemark[4] %
  }
  \\
  {\rm
    Gabi~Dreo~Rodosek\footnotemark[2]~,~%
    Fabio~Pierazzi\footnotemark[3]~,~%
    Lorenzo~Cavallaro\footnotemark[4]~%
  }\\
{\normalsize \footnotemark[2]~~Universität der Bundeswehr München}\\
{\normalsize \footnotemark[5]~~Imperial College London}\\
{\normalsize \footnotemark[4]~~University College London}\\
{\normalsize \footnotemark[3]~~King's College London}
}

\maketitle 

\begin{abstract} 

Machine learning classifiers are vulnerable to adversarial examples---input-specific perturbations that manipulate models' output. Universal Adversarial Perturbations (UAPs), which identify noisy patterns that generalize across the input space, allow the attacker to greatly scale up the generation of such examples. Although UAPs have been explored in application domains beyond computer vision, little is known about their properties and implications in the specific context of realizable attacks, such as malware, where attackers must satisfy challenging problem-space constraints.

In this paper we explore the challenges and strengths of UAPs in the context of malware classification. We generate sequences of problem-space transformations that induce UAPs in the corresponding feature-space embedding and evaluate their effectiveness across different malware domains. Additionally, we propose adversarial training-based mitigations using knowledge derived from the problem-space transformations, and compare against alternative feature-space defenses. 
Our experiments limit the effectiveness of a white box Android evasion attack to \textasciitilde20\% at the cost of \textasciitilde3\% TPR at 1\% FPR. We additionally show how our method can be adapted to more restrictive domains such as Windows malware. 

We observe that while adversarial training in the feature space must deal with large and often unconstrained regions, UAPs in the problem space identify specific vulnerabilities that allow us to harden a classifier more effectively, shifting the challenges and associated cost of identifying new universal adversarial transformations back to the attacker.

\end{abstract}

\section{Introduction}

Universal Adversarial Perturbations (UAPs)~\cite{uaps2017} are a class of adversarial perturbation in which the same UAP can be applied to many different inputs to induce errors in the machine learning (ML) classifier. UAPs have proven to be very effective for crafting practical and physically realizable attacks in computer vision~\cite{uaps2017, mopuri2018nag, khrulkov2018singular, kenny2019ccs} as well as for NLP tasks~\cite{wallace2019nlpuap}, and audio and speech classification~\cite{neekhara2019speechuap, abdoli2019audiouap}.

However, to the best of our knowledge, the study of realizable universal perturbations in ML-based malware detection has not yet been explored, likely due to the difficulty of modifying real-world software while preserving malicious functionality~\cite{pierazzi2020intriguing}. 
Despite this, UAPs remain a tempting attack tool for adversaries, as attackers naturally gravitate towards using low-effort/high-reward strategies to maximize profit~\cite{felt2011survey, fossi2008symantec}. UAPs enable attackers to cheaply reuse the same collection of precomputed perturbations across different malware in order to evade detection. 
As well as being an attractive prospect for individual malware authors, UAPs are promising for the \emph{Malware-as-a-Service} (MaaS) business model~\cite{web:maasKaspersky, web:maasWebroot, web:maasLastline}, in which service providers aim to produce cheap, universally evasive transformations at scale.

In this paper, we analyze the impact of UAP attacks against malware classifiers, revealing that they pose a significant and real threat against ML-based malware detection systems. Firstly, we show that the effectiveness of UAP attacks in the feature space against linear and nonlinear classifiers is comparable to that of input-specific attacks, demonstrating the existence of a systemic vulnerability in these malware detectors. Secondly, our analyses in the problem space confirm this vulnerability. In the process, we propose a methodology to produce functional (real) adversarial malware that rely on UAPs. Specifically, we propose a greedy algorithm that identifies a short sequence of problem-space transformations that, when applied to a malware object, evade detection with high probability while preserving the malicious functionality. 

We provide an extensive experimental evaluation across the Android and Windows malware landscape, exploring linear and nonlinear ML models, including Logistic Regression (LR), Support Vector Machines (SVMs), Deep Neural Networks (DNNs), and an improved variation of Gradient Boosting Decision Trees (GBDTs), known as LightGBM (LGBM)~\cite{ke2017lightgbm}. Our results show that unprotected models are brittle and vulnerable to our UAP attacks, even when the attacker's knowledge about the target classifier is limited.

To defend against this threat we propose a novel method to perform adversarial training using evasive examples created in the problem space. Adversarial training~\cite{goodfellow,madry2017towards} has proven to be one of the most promising defense approaches against adversarial examples, but protecting against multiple perturbation types is challenging~\cite{tramer2019adversarial}. This limitation is supported by our experiments which show that feature-space adversarial training is not a sufficient solution against problem-space UAP attacks. Therefore, we propose an adversarial training method with UAPs by learning from the feature-space perturbations induced by the problem-space transformations. Our approach allows us to protect against a set of manipulations used by an attacker to produce adversarial malware, with a small decrease in the detection rate of non-adversarial malware. Consequently, our method reduces the number of evasive samples required to be crafted in the problem space for a defender to perform  adversarial training. 

We note that we do not provide robustness against all possible adversarial ML attacks. Our defense focuses on ``patching'' the pockets of vulnerabilities that allow adversaries to craft realizable attacks using a predefined toolkit of transformations. While defending against \emph{unknown unknowns} remains an open challenge, our methodology can be realistically applied to harden ML-based malware detection models against known vulnerabilities (\ie the set of transformations that attackers rely on to evade detection). This raises the cost of creating evasive malware, as adversaries must either identify a new set of problem-space transformations, or focus on input-specific attacks that may require longer transformation chains, increasing the risk of malware corruption~\cite{labaca-castro2019armed}.

In summary, this paper makes the following contributions: 

\begin{itemize}[noitemsep]
    \item We first demonstrate that ML-based malware classifiers are especially vulnerable to UAP attacks in the feature space, and empirically show that they achieve similar effectiveness compared to input-specific attacks~(\cref{section:feature-space-uaps}). 
    
    \item We then propose a novel attack methodology to find weaknesses in ML-based malware classifiers using UAPs. This methodology allows attackers to modify real malware in the problem space while preserving malicious semantics and plausibility~(\cref{section:evaluation}). We experimentally demonstrate the effectiveness of our approach by generating highly evasive Windows and Android malware variants using UAP attacks.
    
    \item We lastly propose and evaluate a novel defense to mitigate this threat based on adversarial training, using the knowledge from the evasive malware generated with our UAP attack. Our defense raises the cost for attackers and disincentivizes the use of powerful UAPs~(\cref{section:robustness}). %

\end{itemize}

We release our UAP attacks and defenses for malware as a library, \tool, to foster future research. For the purpose of review, we host the anonymized code here: \url{https://bit.ly/getGAME-UP}.

\section{Background}

We introduce major notation and pertinent background on feature-space and problem-space evasion attacks, UAPs, and adversarial training. In particular, we borrow notation from~\citet{biggio2018wild} and~\citet{pierazzi2020intriguing}. 

\subsection{Adversarial Evasion Attacks}
\label{sec:evasion}

In the malware domain, evasion attacks occur when an attacker modifies an object at test time to evade detection. The object can be represented in two ways: \emph{feature-space objects} are the abstract numerical representation fed to the machine learning algorithm whereas \emph{problem-space objects} represent the original input space, \ie real software applications. 

The feature space, label space, and problem space are denoted by $\mathcal{X}$, $\mathcal{Y}$, and $\mathcal{Z}$, respectively. Each input object $z \in \mathcal{Z}$ is associated with a ground-truth label $y \in \mathcal{Y}$. A classifier $g: \mathcal{X} \longrightarrow \mathcal{Y}$ produces a label prediction $\hat{y} = g(x)$. In order to be processed by a classifier, we must use a feature mapping function to convert it to the feature-space representation such that $\varphi: \mathcal{Z} \longrightarrow \mathcal{X} \subseteq \mathbb{R}^n$. In the software domain, the feature mapping function is not invertible nor differentiable, meaning it is not easy to find a problem-space attack with traditional gradient-based methods; moreover, with respect to the feature space, we must take into consideration several additional constraints to generate realistic, inconspicuous problem-space objects that preserve the attacker-defined behavior.

\paragraph{Feature-space attacks} The goal of the adversary is to transform an object $x \in \mathcal{X}$ into an object $x' \in \mathcal{X}$ in which $g(x') = t \in \mathcal{Y}$ where $t \neq y$. Hence, the adversary forces the model $g$ to predict the incorrect class for the object $x'$ . 
In the malware context, we consider the case in which a malicious object is misclassified as benign. 

\paragraph{Feature-space constraints} A set of constraints $\Omega$ defining possible transformations in the feature-space. For example, limiting the lower and upper bounds of the perturbation 
or the total number of modifiable features.

\paragraph{Problem-space attacks} The goal of the adversary in the problem-space is to find a sequence $\seqT : T_{n} \circ T_{n-1} \circ ... \circ T_{1}$ where each transformation $T: \mathcal{Z} \longrightarrow \mathcal{Z}$ mutates the object $z$ such that $g(\seqT(z)) = t \in \mathcal{Y}$ where $t \neq y$, while satisfying all problem-space constraints defined by the attacker. 

\paragraph{Problem-space constraints} Problem-space attacks must satisfy additional constraints~\cite{pierazzi2020intriguing}: \emph{available transformations}, \emph{preserved semantics}, \emph{robustness to preprocessing}, and \emph{plausibility}. For example, transformations in the problem space are typically limited to addition, since removal or modification can lead to file corruption. For machine learning classifiers relying on static analysis, this is often achieved by injecting instructions that will not be executed or modifying parameters that do not affect the integrity of the file.

\subsection{Universal Adversarial Perturbations}
UAPs are a class of adversarial perturbations where a single perturbation applied to a large set of inputs produces errors in the target machine learning model for a large fraction of these inputs~\cite{uaps2017}. UAPs reveal systemic vulnerabilities in the target models and expose a significant risk, as they reduce the effort for the attacker to create adversarial examples, enabling practical and realistic attacks across different applications as, for example, in computer vision or object detection~\cite{brown2017adversarial,song2018physical,eykholt2018robust,liu2018dpatch}, perceptual ad-blocking \cite{tramer2019adblocking}, or LiDAR-based object detection~\cite{cao2019adversarial,tu2020physically}. As UAPs find patterns the target models are especially sensitive to, attackers can use UAP attacks to craft successful and very query-efficient black-box attacks~\cite{kenny2019ccs}. So far, realizable UAP attacks have not been explored in the context of machine learning malware classifiers.

In our experiments, we measure the \textit{effectiveness} of UAP attacks in terms of the Universal Evasion Rate (UER), computed over a set of inputs $\mathcal{X}$ and defined as:

\begin{equation}
   \text{UER} = \frac{|\{x \in \mathcal{X}: \arg \max  g(x + \delta) \neq y \in \mathcal{Y}\}|}{|\mathcal{X}|} \label{eq:uer}
\end{equation}

That is, UER denotes the fraction of inputs in $\mathcal{X}$ for which the classifier outputs an error when the UAP $\delta$ is applied.

\subsection{Adversarial Training}

Adversarial training is one of the most successful and promising approaches for defending  against adversarial inputs~\cite{goodfellow,madry2017towards}. It involves training a model using adversarial examples crafted for each class so that the model becomes more robust to these types of inputs. The robustness gained depends on the strength and type of examples generated. \citet{shafahi2020universal} also proposed using UAP adversarial training to defend against UAP attacks in computer vision tasks.

However, adversarial training suffers from some limitations. When using standard adversarial training techniques, such as Projected Gradient Descent (PGD) or multi-step PGD, the cost of generating adversarial examples is very high, making them impractical for large-scale datasets---although some more specialized techniques can be used to alleviate the computational burden~\cite{shafahi2019adversarialTr}. On the other hand, defending against multiple perturbations is challenging~\cite{tramer2019adversarial} and making the model robust to certain perturbations can facilitate evasion attacks that use different  perturbations the defender did not consider during training. 

\section{Feature-Space UAPs for Malware}
\label{section:feature-space-uaps}

In this section, we present a motivational experiment to demonstrate that malware classifiers are especially vulnerable to UAPs crafted in the feature space---that is, without considering the set of problem-space constraints which restrict how the attacker can mutate an input object. Although in a domain such as malware, feature-space attacks may be unrealistic from a practical perspective~\cite{pierazzi2020intriguing}, this analysis exposes the systemic risk of malware classifiers to universal perturbations and the importance of understanding this threat in the problem space, as we describe in subsequent sections. To the best of our knowledge, this is the first study of the impact of UAP attacks %
for malware detection. 

We perform an empirical evaluation of feature-space UAP attacks using two well-known malware datasets: i) \sleipnir~\cite{al2018adversarial} for Windows malware and ii) \drebinnew~\cite{drebin2014,pierazzi2020intriguing} for Android malware. 
\sleipnir consists of 34,995 malicious and 19,696 benign PE files and uses a binary feature space where each feature corresponds to a unique Windows API call, with $1$ and $0$ indicating presence and absence of the call, respectively. Each vector (\ie PE representation) consists of 22,761 API calls. The \drebinnew dataset, also a binary feature space, is presented in detail in~\cref{sec:attack-android}. 

For both datasets we create a random split with 60\% of examples used for training and 40\% for testing. Note that, without loss of generality, here we consider \sleipnir as a Windows representative out of simplicity, given its convenient binary feature space, so that we can have a more direct and clearer comparison between Windows and Android malware. Thus, we can use the same type of constraints to model the attack's strength using the $L_0$ norm. For the problem space analysis, in the remainder of the paper will consider a more comprehensive dataset: \ember~\cite{anderson2018ember}, which also includes continuous features~(\cref{sec:attack-windows}). The use of \ember would have resulted in a more difficult comparison for feature space attacks, as we would need to combine different norms to model the attacker's constraints for both the continuous and the discrete features present in the dataset.

For both datasets we train a \emph{Logistic Regression} (LR) classifier and a \emph{Deep Neural Network} (DNN) with the following architecture: $n_f \times 1,024 \times 512 \times 1$, where $n_f = 22,761$ for \sleipnir and $5,000$ for \drebinnew. For the DNNs we use Leaky ReLU activation functions for the hidden layers (with negative slope equal to 0.1) and a sigmoid activation function for the output layer. We include Dropout to reduce overfitting and use the Adam optimizer~\cite{kingma2015adam} with learning rate equal to $10^{-3}$ for both the LR and the DNN.

\subsection{Input-specific vs UAP attacks}
We test the robustness of the LR and the DNN classifiers against input-specific and UAP attacks under perfect knowledge white-box settings. For the input-specific attacks we use the attack proposed by \citet{grosse2017esorics}, which relies on the recursive computation of the Jacobian, searching at each step for the feature that maximizes the change in output in the desired direction (\ie towards evasion). For the UAP attack we propose a method where we select the most salient features computed by the Jacobian averaged over the malware examples in the test set. We define the attacker's feature-space constraints in terms of the $L_0$-norm, i.e., the number of features that the attacker can modify, exploring values from $L_0 = 1$ to $20$. As in \citet{grosse2017esorics}, we further assume that the attacker can only \textit{add} features, in order to preserve malicious functionality, \ie the attacker can only change features from 0 to 1 but not from 1 to 0. For the UAP attack, the effective change in the number of features that are set to 1 after the attack is at most $L_0$ for each input, \ie some of the features for these inputs may already be set to 1, and thus, the UAP does not change their value.

The computation of the attacks against the LR can be simplified: for the UAP attack, we select the features with the most negative weights, \ie we select the top-$L_0$ features that are most indicative of goodware. For the input-specific attacks, for each input, we also select the top-$L_0$ features that are most indicative of goodware and that have value 0 for that specific element.
    
    \begin{figure}[t!]
    \centering
    \begin{subfigure}[t]{1.0\columnwidth}
    \centering
    \includegraphics[width=0.8 \columnwidth]{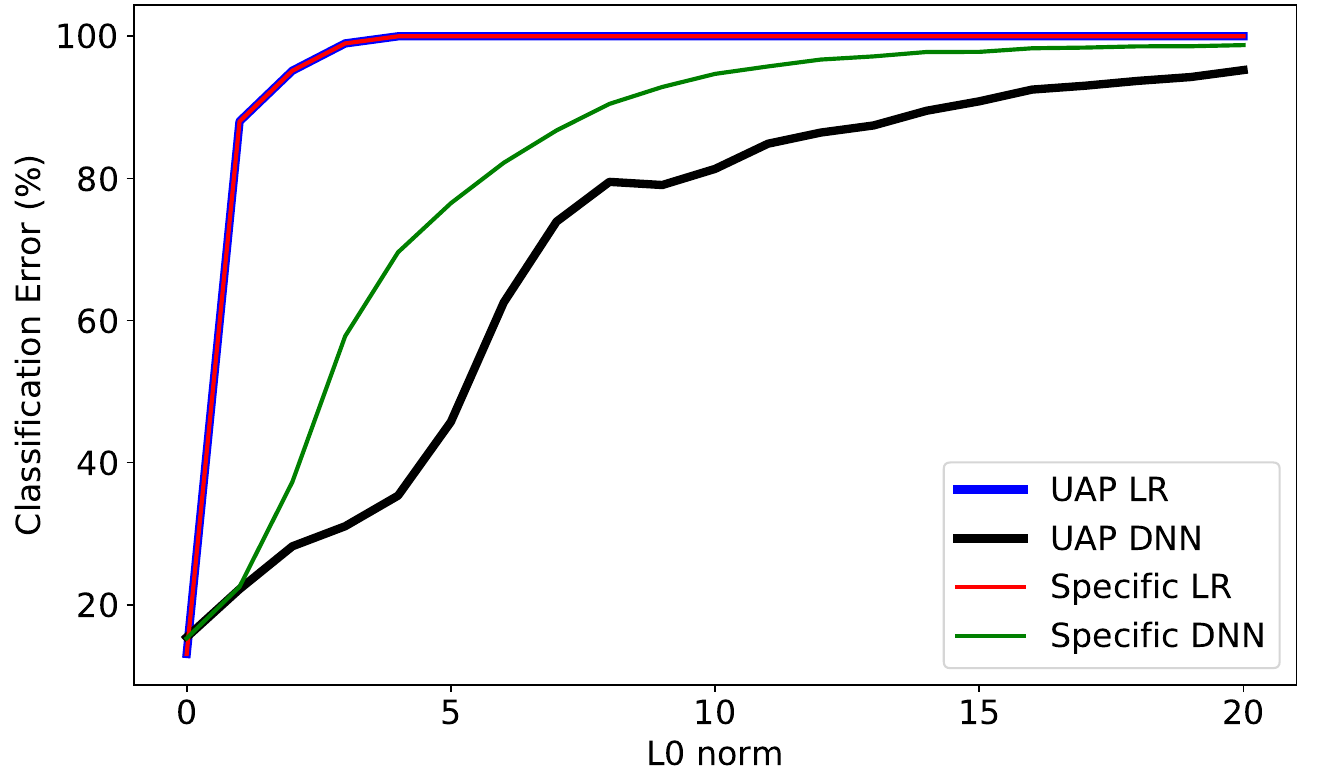} \\
        \caption{Android} %
        \label{fig:stuffb}
    \end{subfigure}
    
    \begin{subfigure}[t]{1.0\columnwidth}
    \centering
    \includegraphics[width=0.8 \columnwidth]{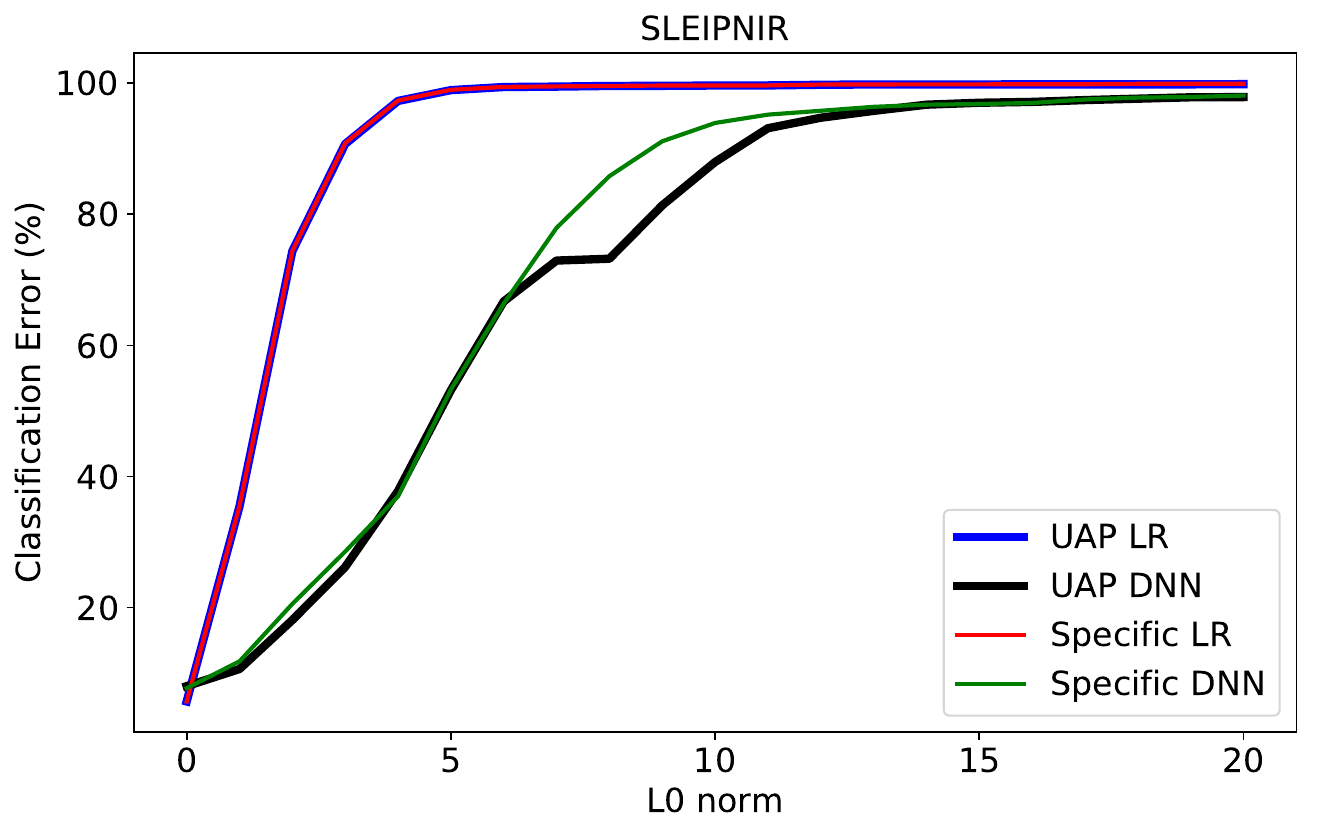} \\
        \caption{Windows} %
        \label{fig:stuffb}
    \end{subfigure}
    \caption{Input-specific \vs UAP white-box attacks in the feature space against LR and DNN for (a) Android malware (DREBIN) and (b) Windows malware (SLEIPNIR).}
    \label{fig:SpecificvsUAP}
\end{figure}

\Cref{fig:SpecificvsUAP} shows the results for the \drebinnew and \sleipnir datasets, reporting the classification error of the adversarial malware at different attack strengths (including when the malware is not manipulated, \ie $L_0 = 0$). We observe that for $L_0 = 20$, the effectiveness of both UAP and input-specific attacks is above $95\%$ in all cases, achieving in some cases effectiveness close or equal to $100\%$. In other words, just by modifying $0.09\%$ and $0.4\%$ of the features used by \sleipnir and \drebinnew classifiers respectively, we can achieve very successful attacks. 

Most importantly, we observe that the effectiveness of the UAP attacks is comparable to those of the input-specific attacks, especially for the linear classifiers, where the results are almost identical. The reason is that, in the case of the LR, the features associated with the most negative weights (\ie those indicative of goodware) are rarely present in the malware examples. Therefore, in most cases, both UAP and input-specific attacks modify the same features.

For lower values of the $L_0$-norm we observe that the DNN is more robust than the LR, and that for \drebinnew, the effectiveness of the UAP attack against the DNN is slightly lower compared to input-specific attacks. However, as previously mentioned, given the very low percentage of features the attacker needs to modify to craft very successful attacks, our results show that both LR and DNN are very brittle and can be easily evaded, which is consistent with previous work~\cite{grosse2017esorics}. 

\subsection{Transferability of UAP attacks}
\label{section:transferability}
\begin{figure}
    \centering
    \begin{subfigure}[t]{1.0\columnwidth}
    \centering
    \includegraphics[width=0.85 \columnwidth]{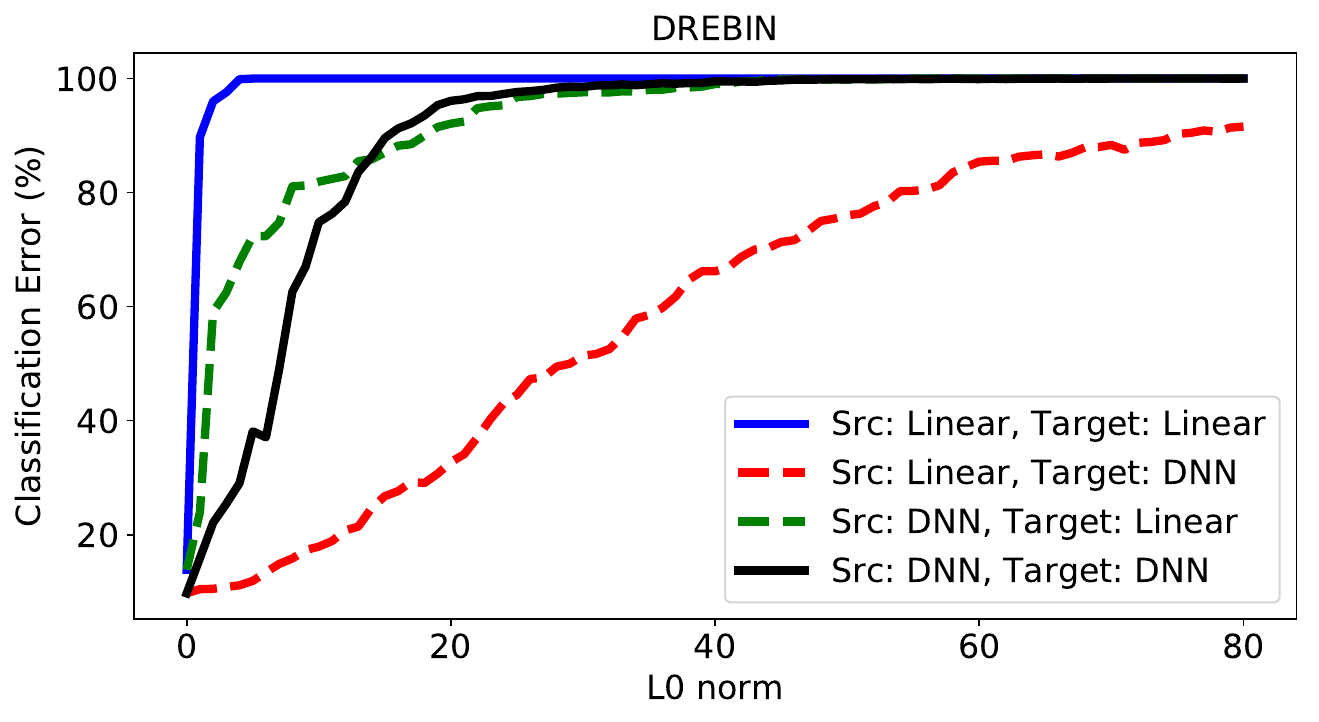} 
        \caption{Android} %
        \label{fig:stuffb}
    \end{subfigure}
    
    \begin{subfigure}[t]{1.0\columnwidth}
    \centering
    \includegraphics[width=0.85 \columnwidth]{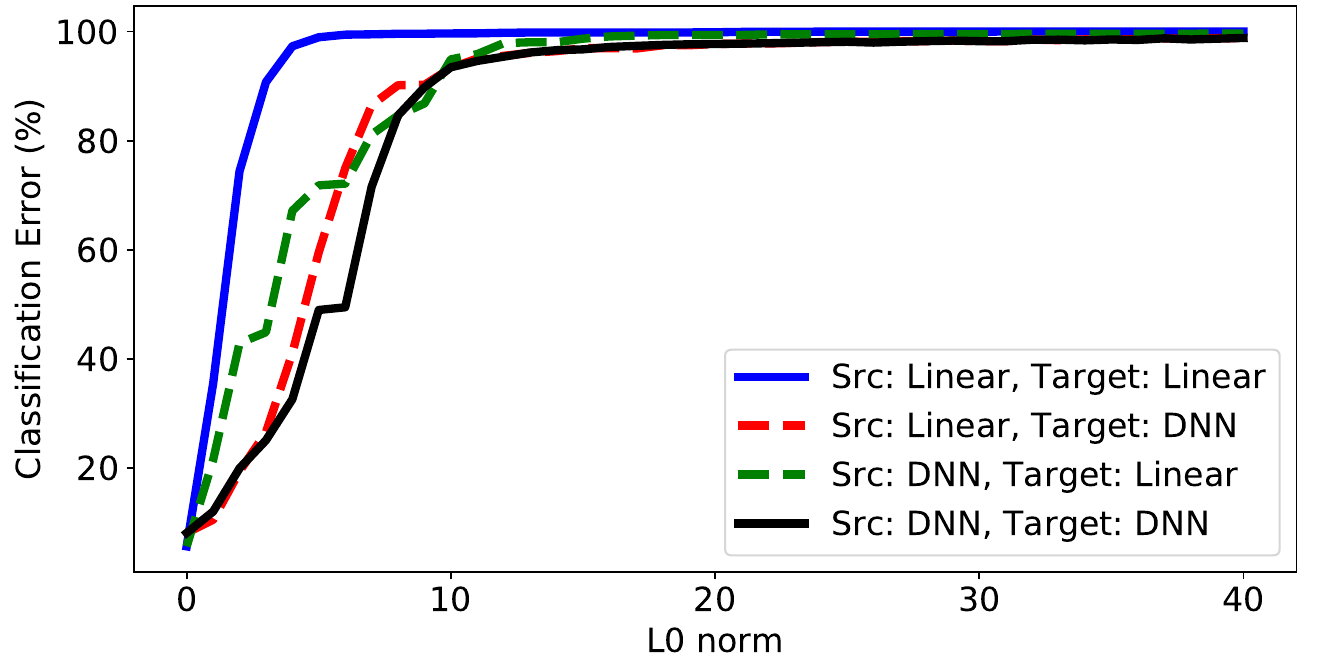} \\
        \caption{Windows} %
        \label{fig:stuffb}
    \end{subfigure}
    \caption{Transfer attacks in the feature space for (a) Android malware (DREBIN) and (b) Windows malware (SLEIPNIR).}
    \label{fig:transferability}
\end{figure}

Additionally we perform an empirical evaluation of the transferability \cite{papernot2016transferability} of UAPs between the linear model and the DNN to better characterize their vulnerabilities in the feature space. Using the same settings as before, we use the UAPs previously generated for both the LR and the DNN and perform transfer attacks. The results are shown in \Cref{fig:transferability} for \drebinnew and \sleipnir. 

For \drebinnew, we can observe that the LR classifier is more brittle to the attacks as compared to the DNN. \Cref{fig:transferability}(a) shows that the white-box attack against the LR achieves a very high effectiveness for very low values of the $L_0$-norm and that the transfer attack using the UAP generated for the DNN is also very effective. Actually, for low values of $L_0$ it is more effective than the white-box attack targeting the DNN itself. On the other hand, we observe that the DNN is more robust to the transfer attack with the UAP generated for the LR and that it requires manipulating \textasciitilde80 features to achieve a success rate greater than 80\%.

For \sleipnir, \Cref{fig:transferability}(b) shows similar results, although in this case, the difference in robustness between the linear classifier and the DNN is not as significant as in the case of \drebinnew. Thus, effectiveness of the transfer attack targeting the DNN with the UAP generated from the LR achieves a high evasion rate for values of $L_0$ close to 30. 
    
\subsection{Discussion}
Our results show the importance and impact of UAP attacks against malware classifiers: they achieve effectiveness comparable to their input-specific counterparts, but pose a significantly higher threat, as the same perturbation generalizes across many malware examples. For unprotected models, the extra capacity of the DNN compared to the linear classifier provides only marginal improvements under white-box settings. However, the transferability analysis shows that the linear classifier is weaker and that the UAP attacks generated with the DNN are highly effective. In Appendix \ref{app:transferabilityDNN} we show additional results demonstrating a very high transferability of the attacks between DNN models with different architectures, further enabling black-box attacks. 

The apparent robustness of the DNN with respect to the linear classifier is not relevant from a practical perspective: as we show in the subsequent sections, when considering practical attacks generated in the problem space, the number of features modified in the feature space as a consequence of the software manipulations can be important. However, when applying defensive techniques such as adversarial training, the DNN offers significant benefits in terms of robustness compared to linear models. %

Our results suggest that systemic vulnerabilities exist in machine learning malware classifiers that attackers can leverage to craft very effective UAPs capable of evading detection, regardless of the malware they are applied to. This reduces the cost for the attacker to generate adversarial malware examples at scale. These results justify the attack methodology considered in the following sections, where we show that it is also possible to generate very effective realizable UAP attacks in the problem space, which pose a significant and real threat.

\section{Problem-Space UAPs for Malware} 
\label{section:evaluation}

Motivated by the results of feature-space UAP attacks in~\cref{section:feature-space-uaps}, next we explore the feasibility of generating \textit{problem-space} UAPs to realize real-world evasive malware. 

While recent work has shown that feature-space UAPs can be employed in attacks at training time, such as backdoor poisoning attacks~\cite{zhang2020uappoison}, here we focus solely on the test phase of the machine learning pipeline. Specifically, we focus on \textit{evasion attacks} (\cref{sec:evasion}) in which the attacker modifies objects at test time to induce targeted misclassifications. 
We envision a profit-motivated adversary such as a Malware-as-a-Service (MaaS) provider~\cite{web:maasKaspersky, web:maasLastline, web:maasWebroot} with two objectives: 

\begin{description}[noitemsep]
    \item[O1.] They aim to \textit{maximize} the amount of malware that can be made undetectable, increasing revenue.  
    \item[O2.] They aim to \textit{minimize} the cost of making a single malware undetectable, reducing expenditure.  
\end{description}

From these objectives it is clear why UAPs are a natural choice: UAPs are \textit{scalable}, amortizing the cost of generating a perturbation over the total number of evasive malware that it produces. To quantify the success of these objectives, we use the \textit{Universal Evasion Rate} (UER) to measure the universality of each perturbation as defined in~\Cref{eq:uer}.

To this end, we describe a generic methodology for generating malware UAPs in \cref{section:methodology}, and then consider two experimental settings across different malware domains. Firstly in \cref{sec:attack-android} we consider an attack against an Android classifier in which the attacker is relatively unconstrained, secondly in \cref{sec:attack-windows},  we consider an attack against a Windows malware classifier in which the attacker is more constrained, with limited knowledge of the target model and a more opaque set of available transformations---typically a MaaS provider would only have access to binaries where it may be difficult even to discern symbols and sections~\cite{patrickevans2020punstrip}.

These different settings help us explore the nuances of physically transforming binaries with UAPs, as well as helping us gauge how realistic the threat of UAPs really are---across different domains.

\subsection{Methodology for Generating UAPs}
\label{section:methodology}

Generating UAPs that can be used with real-world malware is significantly more challenging than generating UAPs in the feature space~(\cref{section:feature-space-uaps}). In order for a UAP to be successfully applied to real-world malware, there must exist some inverse mapping from the UAP feature vector back to the \textit{problem space}; \ie there must exist some chain of real-world transformations which is capable of inducing the feature-space change in the chosen object. While the complexity of software means that how these real-world transformation chains are found is largely specific to the given domain, here we outline a number of common components that make up our overall methodology.

\paragraph{Available transformations} Each domain is initially constrained by the  \textit{available transformations}~\citep{pierazzi2020intriguing}. These represent the specific \textit{toolbox} that the attacker has access to, \eg a set of gadgets to inject~(\cref{sec:attack-android}) or a tool for performing binary mutations~(\cref{sec:attack-windows}). Formally we define it as a set of domain-specific problem-space transformations where each transformation is a function $T \colon Z \rightarrow Z$ that mutates a problem-space object $z \in Z$ into $z' \in Z$. This set is analogous to an \textit{action space} in reinforcement learning (RL)~\cite{wang16dueling}. Generally we assume the attacker is able to add, remove, or modify features arbitrarily, so long as the resulting perturbations correspond to a realizable, functioning input object. However, we do not assume the attacker has access to the original source code, as they may be a third party operating on behalf of the malware author (\eg a MaaS provider). We do not put hard limits on the size of the perturbation in terms of $L_p$ norm, as these have been shown to be inappropriate for formulating problem-space attacks~\cite{pierazzi2020intriguing}---however, we note that larger perturbations often correspond to larger transformation sequences which increase the risk of corrupting the input malware.

\paragraph{UAP search} Next, we perform a \emph{greedy search} for a chain of transformations $\seqT = T_n \circ T_{n-1} \circ \ldots \circ T_1$ which can be universally applied to a set of true positive malware in order to flip their labels to benign---this chain is the problem-space equivalent of a UAP. 
The chain is constructed such that each new transformation aims to maximize UER, however whether this search can be feature/gradient-driven or problem-driven depends on the set of transformations itself.    
In order to avoid experimental bias (\eg data snooping) we conduct this search on an \textit{exploration set}, a partition of the training data~\cite{arp2020dodo}. 
This also simulates our MaaS scenario in which an adversary is interested in \textit{reusing} UAPs on future examples which they may not yet have access to.
Note that we avoid splitting the dataset temporally~\cite{pendlebury2019tesseract} in order to evaluate the attacks in the \textit{absence of concept drift}, as performance degradation induced by the evolution of malware over time may lead us to overestimate the UAP success rate.

\paragraph{Feature space analysis} Finally we evaluate the effectiveness of the discovered UAPs on a separate test set in terms of the UER. 
To understand the effect that the UAPs have on the target classifier---and better understand systemic weaknesses in the model---we analyze the feature-space perturbations induced by the problem-space UAPs.

\subsection{Android Malware UAP Attack}
\label{sec:attack-android}

In the Android ecosystem the attacker is typically less constrained in terms of transformations, as they have access to bytecode with which they can perform more detailed injections at scale. Due to this, here we devise a strong problem-space UAP attack against an Android malware detector.

\paragraph{Target Classifier} For this attack we consider \drebin~\citep{drebin2014}, an Android malware detector which can achieve state-of-the-art performance in the presence of concept drift if retrained with incremental retraining~\cite{pendlebury2019tesseract}. \drebin~\citep{drebin2014} uses a linear Support Vector Machine (SVM) as the underlying classifier. For the SVM regularization hyperparameter we use $C=1$.

\paragraph{Dataset} We adopt the \drebinnew Android malware dataset by~\citet{pierazzi2020intriguing} which consists of 152,632 benign and 17,625 malicious apps from \androzoo~\cite{allix2016androzoo}, following the guidelines of \tesseract~\cite{pendlebury2019tesseract} to avoid spatial bias.
The apps are dated between Jan 2017 and Dec 2018 inclusive. 
The apps are embedded in the \drebin~\cite{drebin2014} feature space abstraction, \ie a binary feature space in which Android components (activities, permissions, URLs, services, etc) are represented as either present or absent. 
The apps have been labeled using a common  criteria~\cite{miller16dimva,pendlebury2019tesseract} in which apps are labeled as \textit{malicious} if they are detected by 4+ VirusTotal AV engines and \textit{benign} if they are completely undetected.\footnote{We note that while the original labeling criteria~\cite{pierazzi2020intriguing, anderson2018ember} discard `difficult to classify' \textit{grayware} with between 1 and 3 (Android) and 1 and 39 (Windows) VirusTotal AV detections, which could result in sampling bias~\cite{arp2020dodo}, this would only be to the advantage of the classifiers under attack (\ie it is harder for an attacker to evade this classifier). This is also true for the potential spatial bias~\citep{pendlebury2019tesseract} present in the original \ember dataset~\cite{anderson2018ember}. 
}

\paragraph{Available Transformations} We adapt the procedure from~\citet{pierazzi2020intriguing} which builds on \textit{automated software transplantation}~\cite{barr2015transplantation}: code gadgets are first extracted from a corpus of benign apps and then injected into a host malware until evasion occurs. Gadgets are extracted recursively to preserve dependencies up to a certain distance to improve plausibility. Although this induces \textit{side-effect features}---extra features which may help or harm the evasion effort---it ensures that the injected gadgets are less conspicuous than, for example, no-op API calls~\citep{rosenberg2018raid} or unused permissions~\citep{grosse2017esorics}. We extract 1,395 problem-space gadgets, based on features considered important with respect to benign examples in our exploration set, to obtain the final set of available transformations $\mathcal{T} = \{\,t_0, \dots, t_{1394}\,\}$, where $t_i$ denotes the injection of gadget $i$ into a given malware. Note that none of the transformations are capable of removal, only addition (\ie setting a feature value to 1).

\subsubsection{Target Model Baseline}
\label{sec:subsec:android-split}

To train the target \drebin model on \drebinnew we use a random stratified split with 33\% hold out, partitioning the dataset into 101,596 and 50,041 examples for training and test, respectively. As we aim to to discover UAPs which are effective against the test data without overfitting, we further divide the training set to use 90\% of the examples (91,436) for the actual training and 10\% (10,160) as the \textit{exploration set}, set aside for the UAP search. As our \textit{adversarial test set}, we consider all true positive malware examples detected by the trained classifier (4,503 examples). On the non-adversarial (clean) test data the model achieves an AUC-ROC of 0.981 and 0.855 TPR at 1\% FPR. 

\subsubsection{UAP Search}

In~\citet{pierazzi2020intriguing}, gadgets are selected greedily based on their total benign contribution (\ie considering side effects) and added until the decision score of the host malware is sufficiently benign. Here we alter the search strategy to consider the UER across all true positive malware examples. We iteratively apply all possible transformations, at each step selecting the one maximizing UER across all true positive malware in the \emph{exploration set}, until either the maximum length for the transformation sequence $\seqT$ is reached, 100\% UER is reached, or no remaining transformations can increase UER. We observe that a maximum sequence length of ten is sufficient. Despite its simplicity, we find this greedy strategy to be very effective at searching for successful UAPs at a low computational cost. The use of more advanced,  computationally demanding search strategies (\eg genetic algorithms), would be impractical given the large number of transformations available (1,395). We assume white-box access to the model, although the attack can be estimated using local surrogate models. In \cref{sec:lk-android} we demonstrate a completely black-box alternative with comparable attack success.

\subsubsection{Results Analysis}

\begin{figure}[t]
    \centering
    \includegraphics[width=0.8\columnwidth]{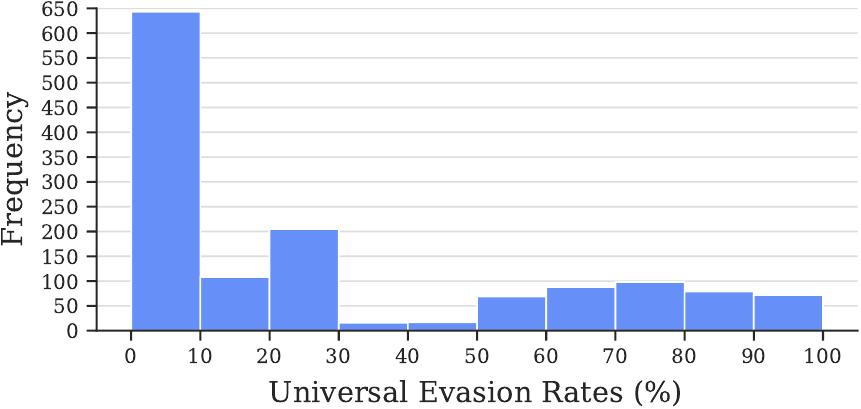}
     
    \caption{Histogram of Universal Evasion Rates (UERs) induced by each available \emph{individual} problem-space transformation targeting the linear \drebin Android malware detector.}
    \label{fig:single-dist-test-linear}
\end{figure}

The strongest UAP that we discover using the exploration set produces 4,413 evasive variants on the test set after a single transformation (98\% UER) and achieves 100\% UER after only two transformations.
As the attack seems very strong, achieving 100\% UER long before the maximum chain length of 10 is reached during exploration, we next investigate the strength of each transformation individually, as shown in~\Cref{fig:single-dist-test-linear}. While 46\% of the transformations achieve less than 10\% UER, 29\% achieve UER of 50\% or greater, with 5\% of the transformations being at least 90\% effective. 

\begin{figure}[t]
    \centering
    \includegraphics[width=\columnwidth]{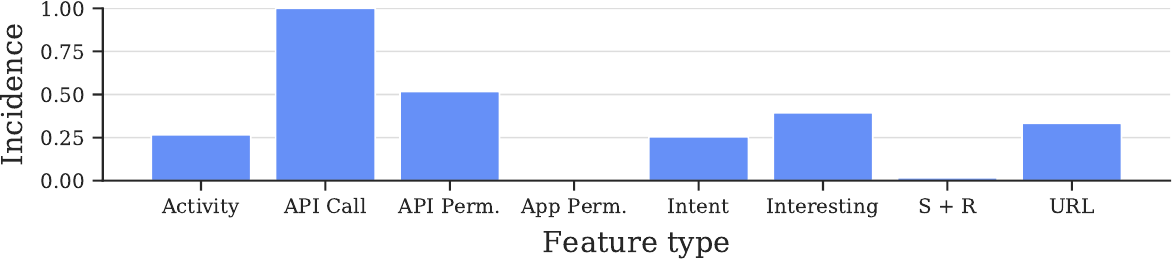}
    \caption{Relative incidence of feature perturbations, grouped by type, induced by the most effective individual transformations (UER  $\geq 90\%$) targeting  \drebin.}
    \label{fig:single-dist-relinc-top}
\end{figure}

We next examine the nature of the feature-space perturbations induced by these strong transformations, to better understand the weaknesses of the classifier. \Cref{fig:single-dist-relinc-top} shows the relative incidence of features, grouped by feature type, across the highly effective transformations (\ie with  UER $\geq 90\%$). The most common feature types perturbed by the UAPs are related to API calls, with API calls perturbed by all transformations, API-related permissions perturbed by half, and a special category of ``interesting'' API calls being the third most common. However, the \textit{individual} features which occur consistently across all of the top transformations are Activities, such as \texttt{activities::CloudAndWifiBaseActivity} (which is present in all but two of these transformations). 

Although we reiterate that $L_p$ norm constraints on the perturbation are not appropriate for problem-space attacks as the object can be modified arbitrarily so long as the problem-space constraints are not violated~\cite{pierazzi2020intriguing}, it is still worth examining the size of the $L_0$ distortion induced by each transformation given how strong they appear to be individually. 

\begin{figure}[t]
    \centering
    \includegraphics[width=\columnwidth]{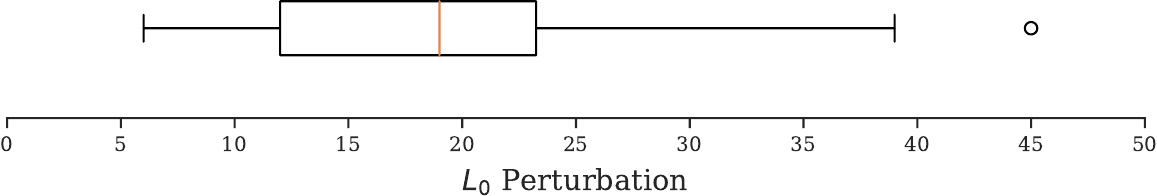}
    \caption{Distribution of $L_0$ norm perturbations (\ie number of changed binary features) induced by the most effective individual transformations (UER  $\geq 90\%$) targeting \drebin.}
    \label{fig:single-dist-l0-top}
\end{figure}

\Cref{fig:single-dist-l0-top} shows the distribution of $L_0$ perturbation sizes, with a mean and median of 18.5 and 19, respectively. To provide perspective, the $L_0$ perturbation induced by the strongest transformation chain is 19; the mean and median $L_0$ norms of the \drebinnew dataset are 50 and 49, respectively. 

\subsubsection{Limited Knowledge Variation}
\label{sec:lk-android}

\begin{figure}[t]
    \centering
    \includegraphics[width=\columnwidth]{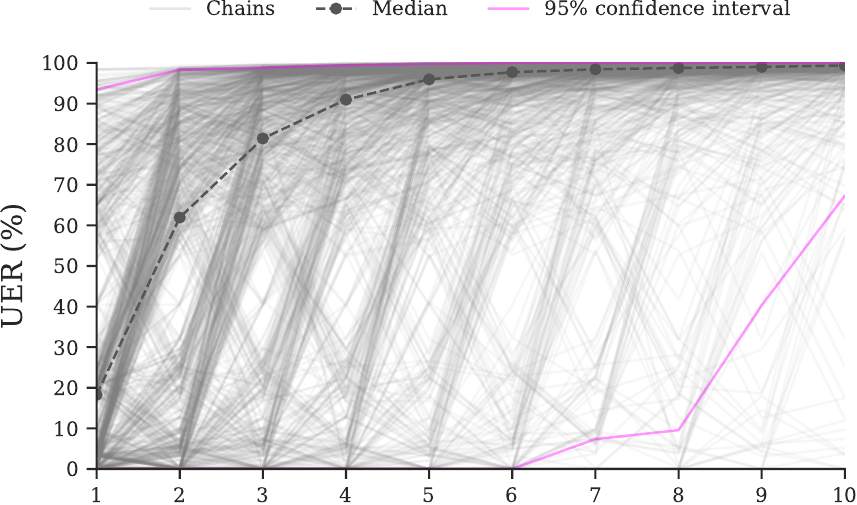}

    \caption{Limited Knowledge (LK) attack against linear \drebin Android malware classifier. Universal Evasion Rates (UER) for 1000 random transformation chains up to length 10. Relatively few transformations are required to achieve a high UER, highlighted by the median at each stage.}
    \label{fig:random-uer-test-linear}   
\end{figure}

\Cref{fig:random-uer-test-linear} shows the results from a na\"ive black-box attack in which $\seqT$ is constructed by selecting gadgets at random without requiring knowledge of each gadget's benign contribution. Each line depicts the UER produced by one of 1,000 transformation chains, tested at each stage of construction. The attack still appears to be extremely potent, with chains at length 5 achieving a median UER above 90\%. 

While clearly effective, the Android domain is naturally more amenable to powerful attacks. The attacker is able to directly manipulate the bytecode, with established program analysis tools such as Soot~\citep{soot99valleerai} and FlowDroid~\citep{arzt14flowdroid} making specific alterations relatively straightforward. Additionally, the toolkit of transformations in the Android attack simplifies the search, as the UER is monotonic with respect to gadget injection---there is no risk of a transformation reducing the evasiveness of the transformation chain.

\subsection{Windows Malware UAP Attack}
\label{sec:attack-windows}

Having demonstrated the impact of problem-space UAPs for Android, here we explore how effective UAPs can be in the Windows domain where the attacker is more constrained.%

Transforming Windows PE binaries is more challenging as they are more prone to corruption during problem-space transformation than Android apps, due to lack of access to source code or bytecode. For example, the effect that an individual transformation (\eg `UPX pack') will have on the input binary, and thus, in the feature space, cannot be calculated a priori. Because of this fragility, a common semantic-preserving attack is to simply append random bytes to the end of the binary~\cite{kolosnjaji18byteappend,web:cylance}. However, these transformations may be detected and removed before classification. Conversely, using more sophisticated transformations increases the risk of disrupting the original malicious semantics and transformations that subtract malicious features (such as packing or compression) may equally obfuscate benign features. In the remainder, we adapt a variety of Windows problem-space transformations from related work, and use them to build problem-space UAP attacks.

\paragraph{Target Classifier} For this attack we consider a state-of-the-art Windows malware detector proposed by~\citet{anderson2018ember} which uses a LightGBM model~\cite{ke2017lightgbm} and default hyperparameters of 100 trees with 31 leaves each.

\paragraph{Dataset} We focus on the \ember Windows malware dataset by~\citet{anderson2018ember} which consists of features extracted from 400,000 benign and 400,000 malicious PE files (as well as 300,000 unlabeled examples which we discard). The remaining apps are mostly dated between Jan 2017 and Dec 2017 inclusive with \textasciitilde4\% predating 2017. The examples have been labeled as \textit{malicious} if they are detected by 40+ VirusTotal AV engines and \textit{benign} if they are completely undetected. The \ember feature space has three broad types of features related to parsed features (\eg file size, header information), format-agnostic histograms (\eg byte-value/entropy histograms), and printable strings (\eg character histograms, average length, URL frequency). Note that unlike \drebinnew, \ember includes continuous features.

For the UAP search we defined two datasets with 1,100 binaries each classified as malicious by the target model, to which we can apply problem-space transformations. The files have been collected from public repositories~\citet{web:virusshare, web:virustotal}. %

\paragraph{Available Transformations} We use the transformations proposed by~\citet{anderson2018learning} and implemented by~\citet{labaca-castro2019aimed}. The transformations are byte-level modifications which can be divided into three categories; \textit{i) inclusion}: adding a new unused section, appending bytes with random length to the space at end of sections or end of the file, adding unused functions to the import address table (similar to~\citet{hu2017generating}); \textit{ii) modification}: renaming sections using alternatives parsed from benign files, manipulating the checksum, debug, and signer info in the header; \textit{iii) compression}: packing or unpacking files using UPX~\cite{web:upx} with random compression rates. \Cref{tab:windows-transforms} summarizes the complete set of transformations.

\begin{table}[t]
\centering
\footnotesize
\caption{Set of available transformations to inject on Windows PEs to target EMBER-LGBM classifier.}
\begin{tabular}{llll}
  \toprule 
  $t_i$ & {\sc Explanation} & $t_i$ & {\sc Explanation} \\
  \midrule 
  $t_0$ & Append overlay & $t_5$ & Remove signature      \\
  $t_1$ & Append imports & $t_6$ & Remove debug          \\
  $t_2$ & Rename section & $t_7$ & UPX pack              \\
  $t_3$ & Add section & $t_8$ & UPX unpack               \\
  $t_4$ & Append section & $t_9$ & Break optional header \\
  \bottomrule 
\end{tabular}
\label{tab:windows-transforms}
\end{table}

\subsubsection{Target Model Baseline} 
\label{sec:subsec:windows-split}
The target model is trained on the \ember~\cite{anderson2018ember} dataset using 300,000 malicious and 300,000 benign examples. The remaining 100,000 malicious and 100,000 benign examples comprise the clean (non-adversarial) test set. As the original dataset contains only extracted features, we augment the dataset with 1,100 malicious binaries downloaded from the VirusTotal~\cite{web:virustotal} and VirusShare~\cite{web:virusshare} repositories, to facilitate the problem-space attacks. All of these samples are successfully detected by the trained model. This set is partitioned in two to obtain an \textit{exploration set} of 100 samples used to search for UAPs, and an \textit{adversarial test set} of 1,000 samples used to validate their effectiveness. On the clean (non-adversarial) test data the model achieves an AUC-ROC of 0.999, with 0.921 TPR at 0.1\% FPR.

\subsubsection{UAP Search}

To search for problem-space UAPs, we apply each of the ten transformations to each of the samples in the exploration set. %
To maximize effectiveness of the UAP, we observe that a single transformation is unlikely to result in a large number of evasions (compared to the strength of a single injection in the Android setting), so relying on the hard labels output by the classifier to maximize UER directly may not give enough information to guide the search, as UER is likely to be 0\% for the first search iteration. Given this, we assume that the attacker has access to the confidence scores (soft labels) output by the classifier and we average these predictions across the modified examples. The transformation that minimizes the average confidence is selected, and chosen in the first position of the transformation chain. In the subsequent rounds, the same procedure is used to search for the next best transformation. 
While access to the classifier's soft labels may not always be possible, certain commercial classifiers are known for providing confidence levels for their predictions~\cite{web:vtanalysis2022}. On the other hand, our results on attack transferability in~\cref{section:transferability} and~\Cref{app:transferabilityDNN} show that effective transfer attacks are possible. Thus, in LK settings, attackers can train a surrogate model and perform successful attacks by exploiting transferability. Alternatively, given that the number of available transformations is low, the cartesian product of transformations can be computed until a length of chain is reached, at the cost of exponentially increasing computational burden.

The number of search rounds $r$ is a product of the size of the exploration set $E$, desired  transformation chain $\seqT$, and available transformations, $\mathcal{T}$---\ie $r = |E| \cdot |\seqT| \cdot |\mathcal{T}|$. This process continues until the maximum of $|\seqT|=10$ is reached. After the process completes, the adversarial malware is executed in a sandbox to verify it has not been corrupted.  %

We have investigated an additional search strategy using genetic programming, however it gravitates towards favoring repeating transformations, which compromise the plausibility of the adversarial examples. Further results are  presented in~\Cref{tab:uer-windows-gp} in~\Cref{app:extra-gp-result}. 

\subsubsection{Results Analysis}

Here we analyze the effectiveness of the most successful candidate transformation chains after applying them to the adversarial test set. 
We perform experiments using UER at $|\seqT|$ of 1, 4, and 10, but here analyze UAP vectors of length four since no improvements were obtained with longer chains.%

\begin{figure}[t]
    \centering
    \includegraphics[width=0.95\columnwidth]{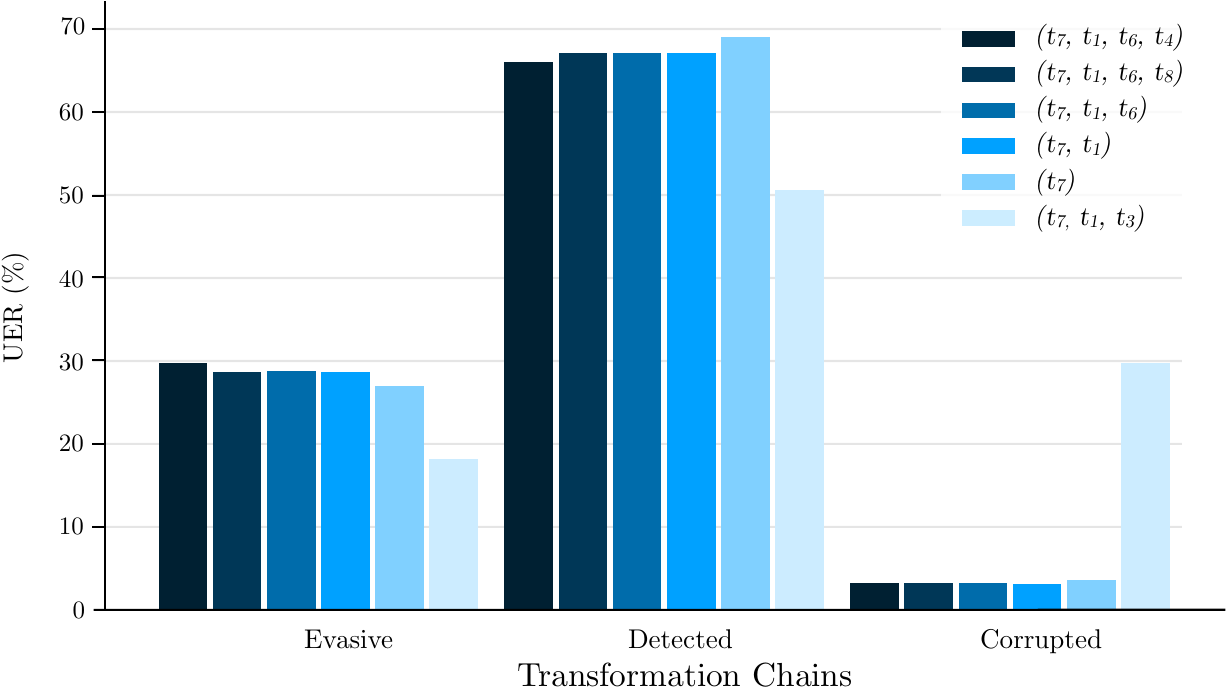}
    \caption{Best candidate UAP chains in the validation set against EMBER-LGBM classifier. The leftmost UAP, $(t_7, t_1, t_6, t_4)$, results in successful adversarial examples for 298 malware whereas $(t_7, t_1, t_3)$ is used as a control to demonstrate how appending specific transformations can drastically increase the corruption rate and limit the success of a chain.}
    \label{fig:uapv_validation}
\end{figure}

We examine six candidates: three of the top scoring chains, two high scoring chains with shorter lengths, and a low scoring chain. \Cref{fig:uapv_validation} shows the success rates for each of the six chains. The most successful chain $(t_7, t_1, t_6, t_4)$ produces 298 evasive variants from 992\footnote{Eight files return 
parsing errors
and have been excluded.} files (30\% UER). Chains $(t_7, t_1, t_6, t_8)$ and $(t_7, t_1, t_6)$ both produce 288 evasive variants each (29\% UER), yet the latter requires only three transformations rather than four. This indicates that in the first two chains, the final transformation does not move the example significantly toward the decision boundary compared to the earlier three. An even shorter chain, $(t_7, t_1)$, also achieves a relatively high evasion rate of 287 evasive variants (29\%) using only two transformations. As the most common initial transformation, a single application of $t_7$ produces 270 evasive variants (27.2\%), which clearly indicates how susceptible the model is to UPX packing. This is likely due to the model paying special attention to the structure of the binary file and therefore being more sensitive to changes caused by high-compression ratio packers. These results support prior work which the distributions of header information for benign and malicious samples packed with UPX to be very similar~\cite{aghakhani2020packin}. 

On the other hand, successful initial transformations do not guarantee a successful chain. For example, appending transformation $t_3$ to the highly effective vector $(t_7, t_1)$ causes the evasion rate to decrease by almost 40\%. This is because the combination $(t_7, t_1, t_3)$ produces almost 10 times more corrupt examples than the alternatives, with 301 non-functional files compared to an average of 33 across other chains.  

The UAP search provides us with very useful insight about how to bypass a classifier in the problem space using a limited transformation toolkit. However, to better understand why some transformations have such a good impact on decreasing the confidence rate of the model, we analyze the feature-space perturbations induced by the chains.

We observe that applying a single transformation results in 27.7\% of features being modified, on average. This number does not increase significantly given longer transformation chains. While this may be caused by the analyzed candidate chains being relatively similar (as the greedy search discards most poor candidates), the most important features for the classifier seem to be perturbed regardless of the chain length. 

\begin{figure}[t]
    \centering
    \includegraphics[width=0.83\columnwidth]{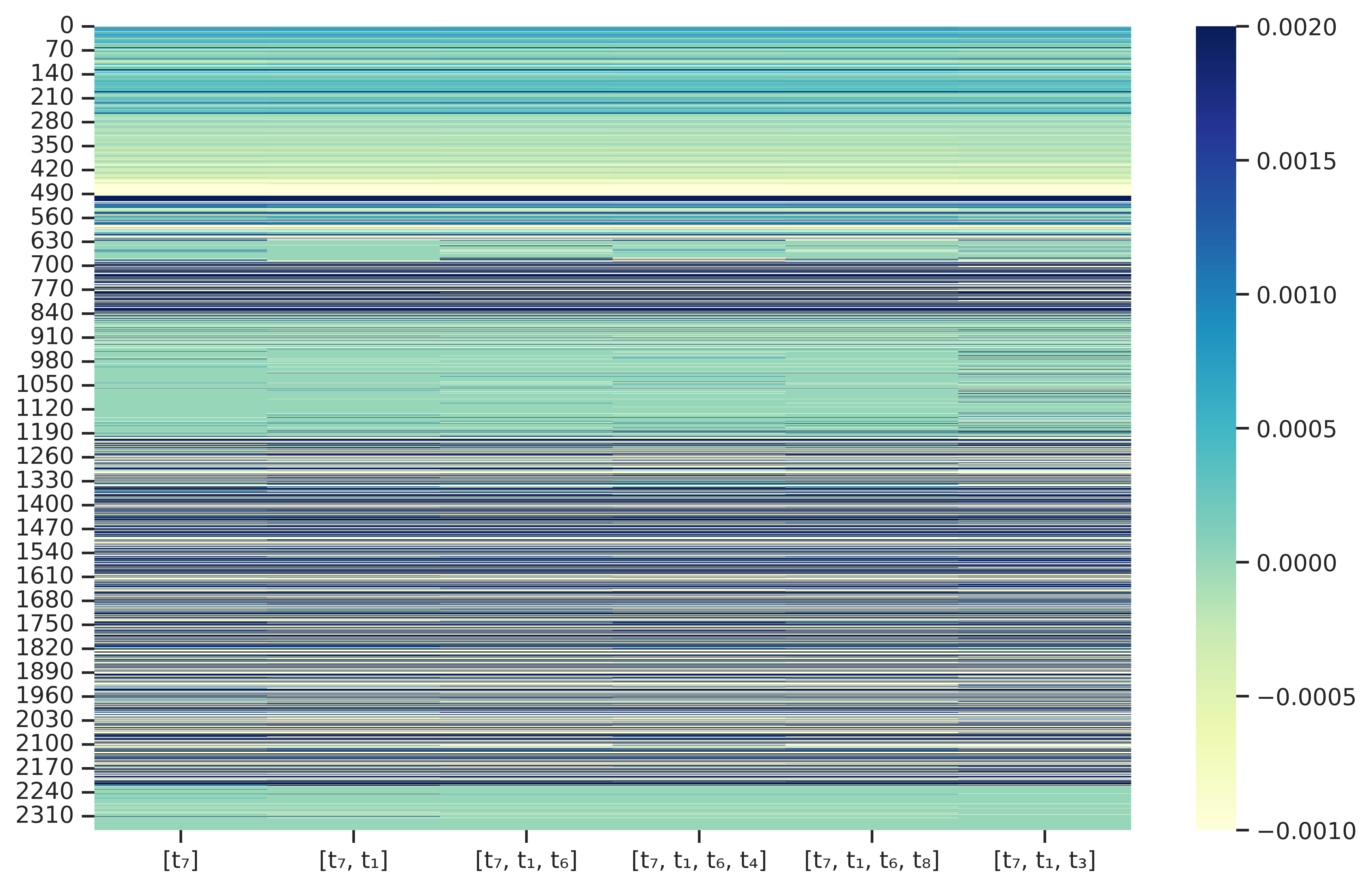}
    \caption{Average delta variation for Windows malware targeting EMBER-LGBM in the exploration set after being injected with each transformation chain. The first five chains (from left to right) are the best candidates for UAP whereas the less effective chain, $(t_7, t_1, t_3)$, shows visible differences within its features. The $y$-axis represents the mapping of the feature-space for each malware example.}
    \label{fig:heatmap_delta_var}
\end{figure}

Furthermore, we analyze the average value of the change (`delta variation') for each feature, for each of the selected candidate chains. As shown in \Cref{fig:heatmap_delta_var}, despite the individual success rates of each candidates, the features with high variation appear to be uniform across all chains. 
However, we do see a distinct difference in the values of features 800 through 1,000 for the less effective candidate $(t_7, t_1, t_3)$. 
This feature group relates to information about the binary's sections, such as names and sizes, which appear to correspond to the final transformation $t_3$, `add section'. 

\section{Evaluating Robustness to UAPs}
\label{section:robustness}

After demonstrating the brittleness of ML malware classifiers against feature- and problem-space UAPs, we now evaluate strategies to improve the resilience of machine learning models against such perturbations. In this section we aim to explore whether we can utilize problem-space knowledge to harden \textit{different types} of classifiers against UAP attacks. 

We introduce our approach to improve defenses, leveraging the UAP attacks in both Android and Windows settings as described in~\Cref{section:evaluation}.
By measuring the effectiveness of new adaptive attacks, after models have been hardened, we determine how each strategy contribute to the robustness of the classifier.

\subsection{Adversarial Training with UAPs}

A promising mitigation against adversarial examples is adversarial training~\cite{goodfellow,madry2017towards,kurakin2016training}: the introduction of evasive examples into the training process to adjust the decision boundary to cover pockets of adversarial space close to legitimate examples.  
However, uniformly applying adversarial training to all regions close to the decision boundary can greatly alter the classifier, such that performance suffers on goodware or previously correctly detected malware. Moreover, effort is wasted in securing regions of the feature space which do not intersect with the feasible problem-space region of realizable attacks~\cite{pierazzi2020intriguing}. 
Our UAPs show that the classifier has specific weaknesses against certain feature types (\eg API calls), so we posit that we can use our UAPs to `patch’ the model against the specific toolkit of available transformations, rather than applying adversarial training indiscriminately. This would significantly raise the bar for attackers, forcing them to obtain a new set of transformations which may not even be possible.

While different ML algorithms necessitate specific adjustments, our process can broadly be defined as follows. 
\emph{i) Generate problem-space UAPs} using a greedy search on the exploration set to calculate the strongest transformation chain, using the toolkit of available transformations, to quantify the model's initial robustness. 
\emph{ii) Adversarially train the model}, either by directly introducing newly generated UAPs to the training process~(\Cref{sec:robustdrebin}) or by using synthetic examples derived from the statistical distribution of examples in the first step~(\Cref{sec:robustember}). 
\emph{iii) Evaluate the robust models} considering an \textit{adaptive attacker}, by performing a fresh search for UAP attacks. We focus on the effectiveness of the UAP attack in terms of UER, and the performance loss incurred on clean data in terms of AUC-ROC and TPR at a fixed FPR of 1\%. 

For how many adversarial examples to include during the fine-tuning we consider two settings. In the \textit{pure} setting, only adversarial examples are used to represent the malicious class. This gives the model the best chance of identifying adversarial inputs, but can cause the TPR to degrade as the model loses the ability to identify clean malware. To mitigate this we also consider an alternative \textit{mixed} setting \cite{kurakin2016training} where clean and adversarial malware examples are interleaved at a ratio of 1:1.

\subsection{Hardening \drebin against UAPs}
\label{sec:robustdrebin}

Here we instantiate our adversarial training-based defense on the Android malware classifiers. 
We hypothesize that the linear \drebin model will not be receptive to adversarial training, as the linear hyperplane will not be flexible enough to adapt to the adversarial inputs, \ie it will begin to `forget' patterns of adversarial inputs seen earlier in the training process~\cite{kirkpatrick2016forgetting}. To test this, we apply our defense to both the linear model from~\Cref{sec:attack-android}, and the nonlinear model originally described in~\Cref{section:feature-space-uaps} which we hereby refer to as \drebin-DNN. Both models are implemented using PyTorch~\cite{paszke2019pytorch}.

We perform the following steps during each of the the last $N$ epochs of the training procedure. At the start of each minibatch, we apply our attack procedure to the partially trained model and search for the most effective UAP transformation sequence, \ie the UAP that maximizes UER across all true positive examples in the minibatch. Next, this UAP is applied to the minibatch malware examples (50\% or 100\% in the mixed and pure settings, respectively).

\paragraph{Results} 
We repeat the white-box attack from~\Cref{sec:attack-android} against both linear and nonlinear models to act as a baseline.\footnote{For completeness, \Cref{app:android-dnn-pk} also demonstrates the vulnerability of the nonlinear model against our limited knowledge attack.} 
We also compare against a number of defenses obtained by generating adversarial examples in the feature space instead of the problem space. These defenses take two parameters: the $L_0$ constraint on the perturbation size and the percentage of adversarial examples to include during the adversarial training procedure. For these we consider $L_0$ constraints of 20 and 40, and mixed/pure adversarial proportions.  

\begin{figure}[t]
    \centering
    \begin{subfigure}[t]{1.0\columnwidth}
        \includegraphics[width=\columnwidth]{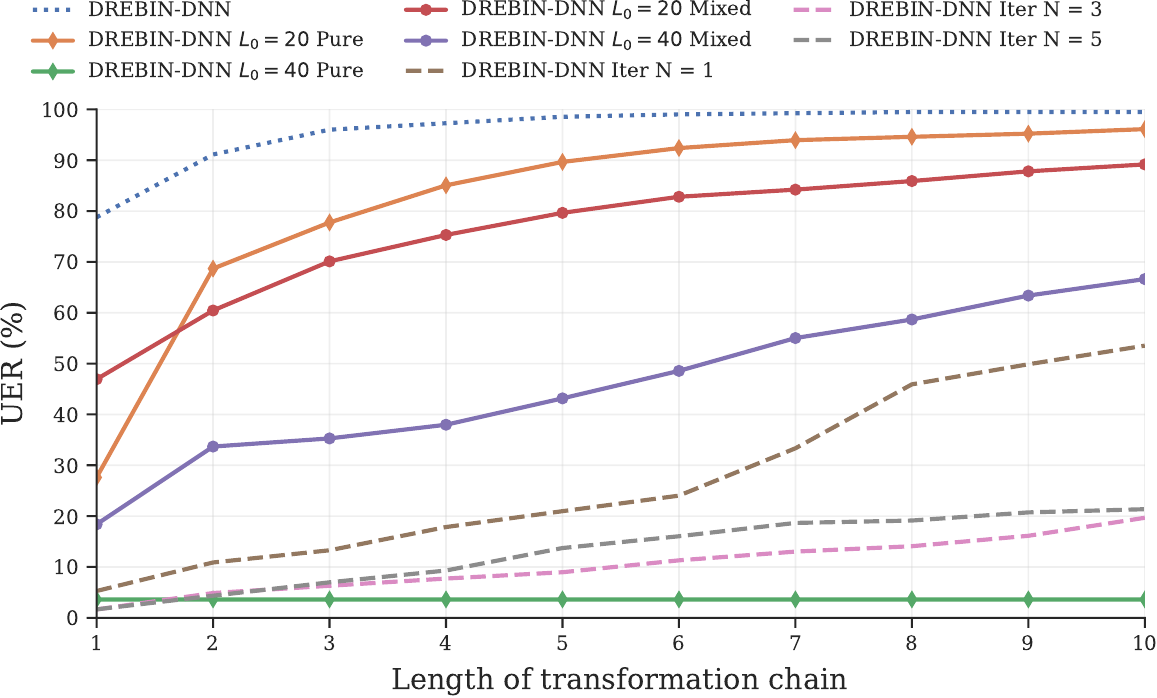}
    
        \caption{\drebin-DNN Android classifiers}
        \label{fig:adv-train-drebin}
    \end{subfigure}
    
    \begin{subfigure}[t]{1.0\columnwidth}
        \includegraphics[width=\columnwidth]{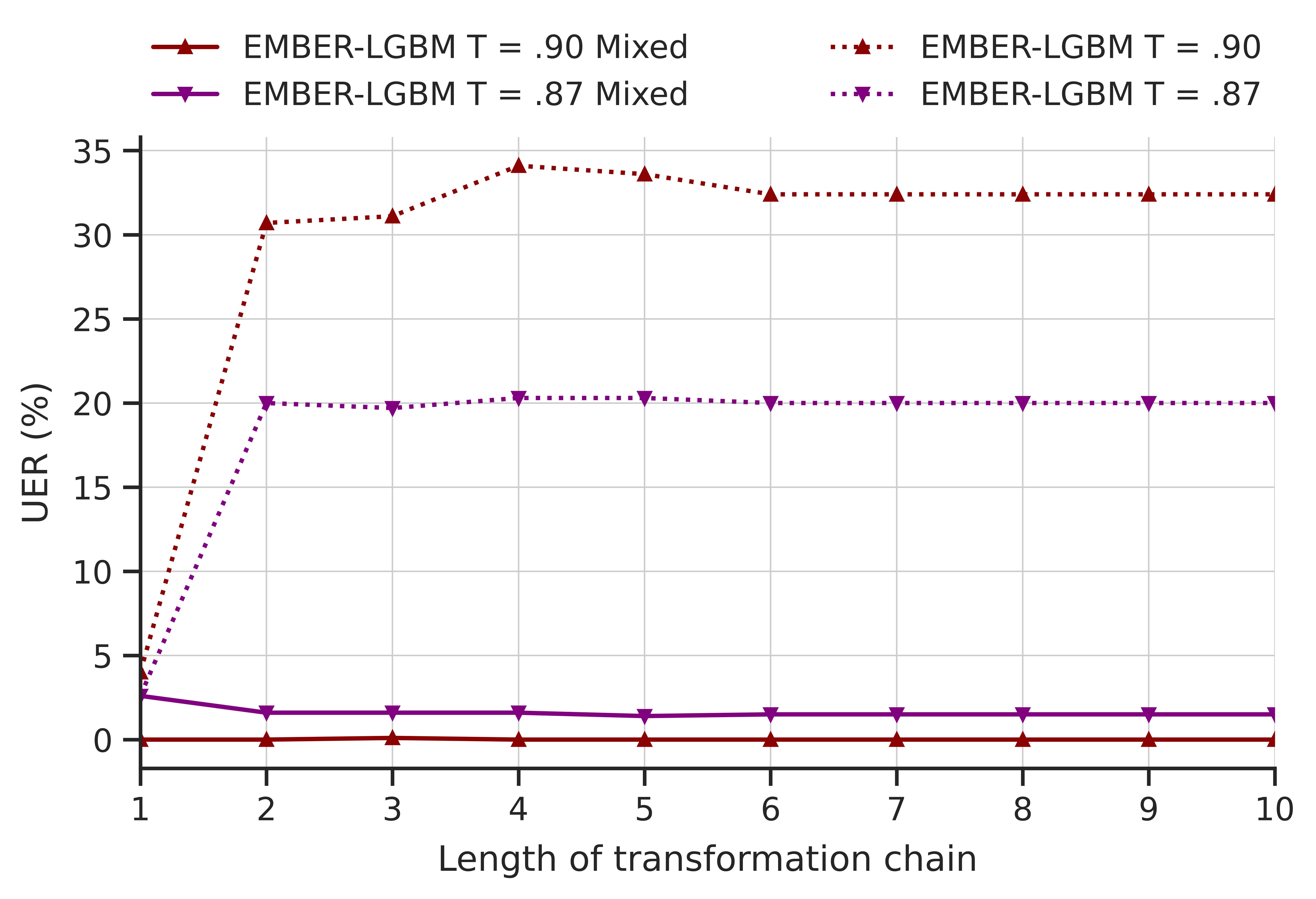}
    
        \caption{\ember-LGBM Windows classifiers}
        \label{fig:adv-train-ember}
    \end{subfigure}
     
    \caption{Adaptive attacks against (a) \drebin-DNN and (b) \ember  classifiers showing increasing Universal Evasion Rates (UER) at varying lengths of problem-space transformation chain. Corresponding performance on clean data is in Tables~\ref{tab:uer-android}--\ref{tab:uer-windows}.}
    \label{fig:adaptive-attacks}
\end{figure}

\Cref{tab:uer-android} shows the results of this procedure applied to the linear \drebin classifier as well as the nonlinear \drebin-DNN, for the last $N = 1$, $3$, and $5$ epochs of training, as well as the UER of the freshly applied adaptive white box attacks (also depicted in \Cref{fig:adv-train-drebin}). 
The close results for $N = 3$ and $N = 5$ suggest an upper bound in the robustness gained, so it is likely that further epochs will result in diminishing returns. 
The results also confirm our hypothesis that the linear model is not as amenable to adversarial training as the nonlinear model. The linear \drebin model shows a larger performance loss on the clean examples compared to the other models (except for $L_0 = 40$ Pure), and while the robustness is improved for small sequences, UER for the sequences of length 10 is $> 80\%$.

Overall, the defense providing the greatest improvement in robustness is $L_0 = 40$ Pure, 
but it comes with a significant performance cost for non-adversarial examples. $L_0 = 40$ Mixed offers a better trade-off with a fairly large increase in robustness without the performance loss. The other feature-space defenses retain their performance on the non-adversarial examples, but do not show a significant gain in robustness. 
However, our approach demonstrates an even greater trade-off than $L_0 = 40$ Mixed, with a similarly small performance loss on clean data, but far greater gains in robustness, reducing the maximum UER of length 10 chains from 99.5\% to \textasciitilde20\%.

\begin{table*}[t]
\centering
\footnotesize
\caption{Comparison of our problem-space defenses against a set of feature-space defenses and undefended models, showing performance on clean examples (AUC-ROC, TPR) and robustness against an adaptive attacker (UER at $|\seqT|$ of 1, 4, and 10).}
\scalebox{1.0}{ %
\begin{tabular}{clrrrrr}
  \toprule 
& {\sc Model} & {\sc AUC-ROC} & {TPR at 1\% FPR} & {\sc $UER_1$} & {\sc $UER_4$} & {\sc $UER_{10}$} \\
  \midrule 
\multirow{2}{*}{Undefended} 
& \drebin                      & 0.981 & 0.855 & 98.7\% &    100\% &    100\%  \\ 
& \drebin-DNN                  & 0.992 & 0.900 & 78.8\% &   97.3\% &   99.5\%  \\
\midrule 
\multirow{4}{*}{Feature-space defenses} 
& \drebin-DNN $L_0 = 20$ Pure  & 0.989 & 0.843 & 27.6\% &   85.1\% &   96.1\%  \\
& \drebin-DNN $L_0 = 40$ Pure  & 0.903 & 0.347 &  3.6\% &    3.6\% &    3.6\%  \\
& \drebin-DNN $L_0 = 20$ Mixed & 0.990 & 0.872 & 46.9\% &   75.3\% &   89.2\%  \\
& \drebin-DNN $L_0 = 40$ Mixed & 0.990 & 0.877 & 18.4\% &   38.0\% &   66.6\%  \\
\midrule 
\multirow{6}{*}{Problem-space defenses} 
& \drebin Iter $N = 1$         & 0.978 & 0.775 & 23.0\% &   70.4\% &   95.7\%  \\
& \drebin Iter $N = 3$         & 0.978 & 0.766 & 21.0\% &   47.0\% &   87.0\%  \\
& \drebin Iter $N = 5$         & 0.978 & 0.761 & 17.4\% &   35.1\% &   82.6\%  \\
\cmidrule{2-7} 
& \drebin-DNN Iter $N = 1$     & 0.990 & 0.874 &  5.3\% &   17.9\% &   53.5\%  \\
& \drebin-DNN Iter $N = 3$     & 0.990 & 0.871 &  1.6\% &    7.7\% &   19.7\%  \\
& \drebin-DNN Iter $N = 5$     & 0.990 & 0.872 &  1.7\% &    9.3\% &   20.4\%  \\
\bottomrule
\end{tabular}}
\label{tab:uer-android}
\end{table*}

\begin{table*}[hbt]
\centering
\footnotesize
\caption{Comparison between baseline (undefended) models and our problem-space defenses against an adaptive attacker (UER at $|\seqT|$ of 1, 4, and 10) using a greedy UAP search strategy. Results using genetic programming are presented in \Cref{app:extra-gp-result}.} %
\scalebox{1.0}{ %
\begin{tabular}{clrrrrrr}
  \toprule 
& {\sc Model} %
& {\sc AUC-ROC} & {TPR} & {\sc $UER_1$} & {\sc $UER_4$} & {\sc $UER_{10}$} \\
  \midrule 
\multirow{2}{*}{Undefended} 
& \ember-LGBM $C = .90$ & 0.999 & 0.921 &   4.0\% &  34.1\% & 32.4\%  \\
& \ember-LGBM $C = .87$ & 0.999 & 0.930 & 2.6\% & 20.3\% & 20.0\%  \\
\midrule 
\multirow{2}{*}{Problem-space defenses}  
& \ember-LGBM $C = .90$ Mixed & 0.988 & 0.836 & 0.0\% & 0.1\% & 0.01\%  \\
& \ember-LGBM $C = .87$ Mixed & 0.998 & 0.853 & 2.6\% & 1.6\% & 1.5\% \\
\bottomrule
\end{tabular}}
\label{tab:uer-windows}
\end{table*}

\subsection{Hardening EMBER against UAPs}
\label{sec:robustember}

We additionally explore how the concept of introducing problem-space information to adversarial training can be adapted and extended beyond neural networks and applied to different machine learning models. Hence, we have adjusted the process to make state-of-the-art classifiers in the Windows domain more resilient to such attacks.

For Windows PE binaries, boosting models such as LGBM have proven to be highly accurate for malware classification in this domain~\cite{anderson2018ember, anderson2018learning}. However, generating adversarial examples in the problem space is significantly more expensive than in the feature space for Windows than it is for Android. Additionally, while the non-linearity of LGBM should be receptive to adversarial training, the model is not trained in batches across multiple epochs as is the DNN. 

To overcome this limitation, we generate an approximation of the feature-space perturbations induced by the problem-space toolkit. For this we create a statistical model where, for each feature, we compute the probability of the feature being modified as a result of the problem-space UAP attack. At training time, we generate adversarial malware in the feature space by sampling random perturbations using this statistical model. 
This allows us to significantly reduce the number of problem-space adversarial objects that need to be generated.
Note that our approximation does not take into account possible interactions between features in the feature space, \ie the statistical model assumes independence across features, and although this is likely not always the case, our empirical results show that even this simple statistical model allows us to harden the ML models against adaptive UAP attacks.

\paragraph{Results} Similar to the Android feature-space in \cref{sec:robustdrebin} we perform adversarial training in Windows using both strategies: pure and mixed. As expected in the problem space, the former model incurs a heavy cost for clean detection performance, with an AUC-ROC of $.624$, while the latter retains better performance at $.853$. Hence, we focus only on mixed. 

As observed in~\Cref{tab:uer-windows} we consider two decision thresholds, $C = .90$ often used in previous work~\cite{anderson2018learning, anderson2018ember, labaca-castro2019aimed} and $C = .87$, corresponding to 0.1\% FPR in our setting. 
We note that it is common practice to use a lower FPR threshold for Windows than in Android \citep[\eg][]{anderson2018learning, anderson2018ember, labaca-castro2019aimed} and that classifiers can leverage FPR as low as $10^{-5}$ to reduce the impact of mistakes~\cite{kaspersky2021whitepaper}.

Following best practices, we perform a fresh adaptive attack against the hardened model. The most effective transformation chains found, $(t_3, t_3, t_0)$ and $(t_3, t_3, t_4, t_8)$, are still successfully detected by the model %
and do not represent meaningful threats. Furthermore, we observe in~\Cref{fig:adv-train-ember} that the model becomes much better at identifying false negatives, as shown by the decrease in UER compared to the baseline. In fact, 99.8\% of the attempts are successfully detected. Despite the capacity of the new model to successfully detect most adversarial examples, compared to the DNN models used for DREBIN, in this case the LGBM offers less flexibility to adapt the UAPs during training, which has a negative impact on the detection of genuine malware, \ie the TPR evaluated on genuine malware decreases.

\section{Discussion}
\label{section:discussion}

\paragraph{Modeling the attacker's constraints} Unlike in computer vision applications, the generation of adversarial examples in the malware domain is subject to specific problem-space constraints that limit how input objects can be modified, to generate realistic working software that preserves malicious functionality~\citep[\eg][]{labaca-castro2019armed, demetrio2020adversarial}. 
Because of this, analyzing feature-space robustness of ML models to certain adversarial examples provides an unrealistic view of the models' vulnerabilities. Many of the attacks available in the feature space are potentially infeasible in the problem space and additionally, modeling appropriate attacker constraints in the feature space can be prohibitively challenging. For example, \ember contains a mixture of discrete and continuous features, making it difficult to model comprehensive constraints (\eg using a combination of $L_p$-norms), even when ignoring some of the problem-space constraints. 

\paragraph{UAP attacks} Our experimental results in \cref{section:evaluation} show that UAP attacks represent an important and practical threat against ML-based malware detectors. However, our results report a disparity on the effectiveness of the attacks for Windows and Android malware. For \drebin we can craft very successful UAP attacks that, in some cases achieve 100\% UER. In contrast, for \ember the UER of the attacks is approximately 30\%. There are two important reasons for this behavior. First, for Windows malware a more limited set of transformations is available to manipulate the malware. Due to the closed-source nature of PE files the transformations need to be designed at a byte-level, which drastically reduces the attack surface for this platform. Secondly, the application of these transformations is more likely to produce  non-functional malware. In contrast, the Android platform allows for a greater number of transformations that can be used to generate adversarial malware, and the addition of these does not have as significant an impact on malware functionality. 
Our results on Android demonstrate the impact of our contribution as the effectiveness of realizable UAP attacks is comparable to that for input-specific attacks, but at a much reduced cost for the attacker, as the same adversarial perturbation can be successfully applied to make many malware evasive. This also evidences a systemic vulnerability in malware detectors. While the effectiveness of the UAP attacks in Windows may appear more modest, this is not necessarily due to the methodology we propose but to the limitations in the tooling to transform PE files. The use of input-specific attacks may not always bring significant improvements in terms of effectiveness: reported evasion rates are typically close to 24\%~\cite{anderson2018learning,labaca-castro2019aimed} and while recent approaches~\cite{labaca2021aimed} may achieve 40\% evasion, the benefit of such improvements may not compensate the extra computational cost of training RL agents and computing fresh attacks individually for many inputs.

\paragraph{Defenses} We show that our methodology for adversarially training the models allows us to harden the model against UAP attacks generated with the considered transformation set. However, we cannot guarantee robustness against other possible transformations that could become available for the attackers, \ie we cannot guarantee robustness against \emph{unknown unknowns}. Compared to adversarial training in the feature space, our methodology focuses on ``patching'' those UAP vulnerabilities that are more relevant from a practical perspective, without having a significant negative impact on the detection of \emph{clean} malware, in particular for DNNs. While we explore limited knowledge attacks in our work, we did not consider \emph{transfer attacks} in the problem space, which can provide a more comprehensive view on the robustness of these models; although the results in the feature space suggest that the susceptibility to UAPs across different models can be similar (\Cref{app:transferabilityDNN}) enabling successful transfer attacks. Also, it would be interesting to analyze if the application of a similar approach can be appropriate for mitigating input-specific attacks. These last two points are left as future work.

\section{Related Work}
\label{section:background}

\paragraph{Adversarial Examples for Malware}
The vulnerabilities of machine learning systems to different threats, both at training and test time, have been investigated for almost 15 years~\cite{barreno2006,huang2011adversarial,biggio2018wild}, attracting a higher attention in the research community since \citet{szegedy2013intriguing} and \citet{biggio2013evasion} showed the existence and weakness of machine learning algorithms to adversarial examples. 
Although the literature in adversarial machine learning has put the focus on computer vision applications, the security community has also started to evaluate the problem on different malware variants, including (but not limited to) Android malware~\cite{pierazzi2020intriguing, grosse2017esorics,demontis2017yes,yang2017malware}, Windows malware~\cite{labaca-castro2019aimed,labaca-castro2019armed,kolosnjaji18byteappend,demetrio2020adversarial,rosenberg2018generic}, PDFs~\cite{vsrndic2013detection,xu2016automatically,biggio2013evasion}, NIDS~\cite{apruzzese2018evading,apruzzese2020deep,corona2013adversarial}, and malicious Javascript~\cite{fass2019hidenoseek}. 
It is important to observe that one peculiarity of the malware domain is that \emph{feature mapping} functions are generally not invertible and not differentiable. This implies that translating an adversarial feature vector in the \emph{feature space} to an actual malware in the \emph{problem space} is significantly more complex. 
To support this challenge, \citet{pierazzi2020intriguing} propose a general framework for problem-space attacks which also clarifies which constraints need to be defined when considering attacks that handle problem-space objects. In our study, we are interested in studying problem-space attacks, and refer to this framework to define our threat model and constraints. 

\paragraph{Universal Adversarial Perturbations} \citet{moosavi2017universal} showed the existence of UAPs, where a single adversarial perturbation applied over a large set of inputs can cause the target model to misclassify a large fraction of those inputs. UAPs expose the systemic vulnerabilities of the model that can be exploited regardless of the input~\cite{kenny2019ccs,ilyas2019adversarial}. UAP attacks are the basis of many physically realizable attacks across different domains, such as image classification~\cite{brown2017adversarial,athalye2018synthesizing,mopuri2018nag, khrulkov2018singular, kenny2019ccs}, object detection~\cite{song2018physical,eykholt2018robust,liu2018dpatch}, perceptual ad-blocking~\cite{tramer2019adblocking}, LiDAR-based object detection ~\cite{cao2019adversarial,tu2020physically}, NLP tasks~\cite{wallace2019nlpuap}, and audio or speech classification~\cite{neekhara2019speechuap, abdoli2019audiouap}.
\citet{hou2020universal} recently explored the use of UAPs in the feature-space of Android malware context. The authors extracted API and hardware information to generate a binary vector that will be modified to attack the model with perfect knowledge. They attack three models and report full evasion only when targeting the \drebin model after nine transformations. The rest two target models, namely DLM and MaMaDroid, required up to 65 and 100 perturbations respectively to achieve close to 100\% rate of adversarial evasions. Compared to them, we report more efficient attacks and a deeper analysis comparing the impact of UAPs and input-specific attacks and analysing the attack's transferability, both for Windows and Android malware.

Different defenses have been proposed to mitigate UAP attacks, most of them only explored in the context of computer vision applications. Some of these defenses aim to detect UAP attacks by trying to denoise the inputs~\cite{akhtar2018defense,borkar2020defending} or, in the case of SentiNet~\cite{chou2020sentinet} aiming to detect adversarial patches in image classification. Other sets of defense aim to harden the model by performing universal adversarial training~\cite{shafahi2020universal,mummadi2019defending}.

\section{Conclusion}
\label{section:conclusion}

After demonstrating that UAPs and input-specific attacks have similar effectiveness in the feature space, we systematically generate and evaluate problem-space adversarial malware using UAPs for both Android and Windows domains. We build on the results to propose a defense: a new variant of adversarial training, also highlighting that nonlinear models such as DNNs are more appropriate than linear classifiers as robust models against problem-space malware UAP attacks. 

\section*{Acknowledgments}
This research has been partially supported by the EC H2020 Project CONCORDIA (GA 830927) and the UK EP/L022710/2 and EP/P009301/1 EPSRC research grants. 

\bibliography{bibliography}

\appendix
\section*{APPENDIX}

\section{Transferability Between DNN Models}
\label{app:transferabilityDNN}
To complete our analysis of the vulnerabilities of the malware classifiers the feature space ({\cref{section:feature-space-uaps}}), we perform an additional experiment on the transferability of UAPs across different DNN model architectures. For this, we use three architectures: \textit{1)} That used for our experiments in \cref{section:feature-space-uaps} (we refer to this model architecture as the \emph{standard model}) with: $n_f \times 1,024 \times 512 \times 1$. \textit{2)} A smaller DNN with the same number of layers as the standard model but with 8-times less units at each layer (we refer to it as the \emph{small model}) with: $n_f \times 256 \times 128 \times 1$. \textit{3)} A deeper DNN model compared to the standard one (we refer to is as the \emph{deep model}) with: $n_f \times 512 \times 256 \times 128 \times 64 \times 1$.

We use the same datasets and settings as in \cref{section:transferability}. We observe that the performance of the three models is very similar in all cases and the model architecture does not have a significant impact on the performance of the malware detector. Thus, for SLEIPNIR the AUC-ROC on the test set is 0.974, 0.974, and 0.973 for the standard, small, and deep model respectively. For DREBIN the test AUC-ROC is 0.989, 0.986, and 0.987 for the standard, small, and deep model. 

\begin{figure}
    \centering
    \includegraphics[width= \columnwidth]{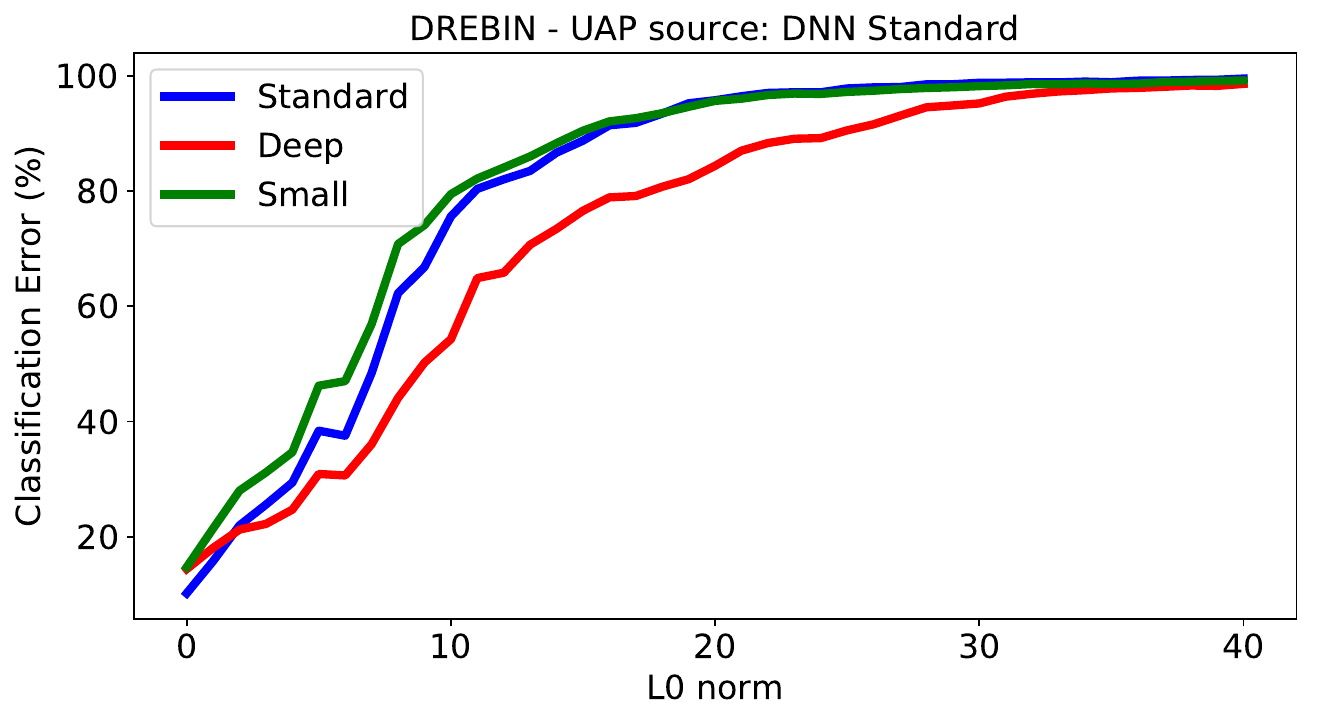} \\
    (a) Android \\
    \includegraphics[width= \columnwidth]{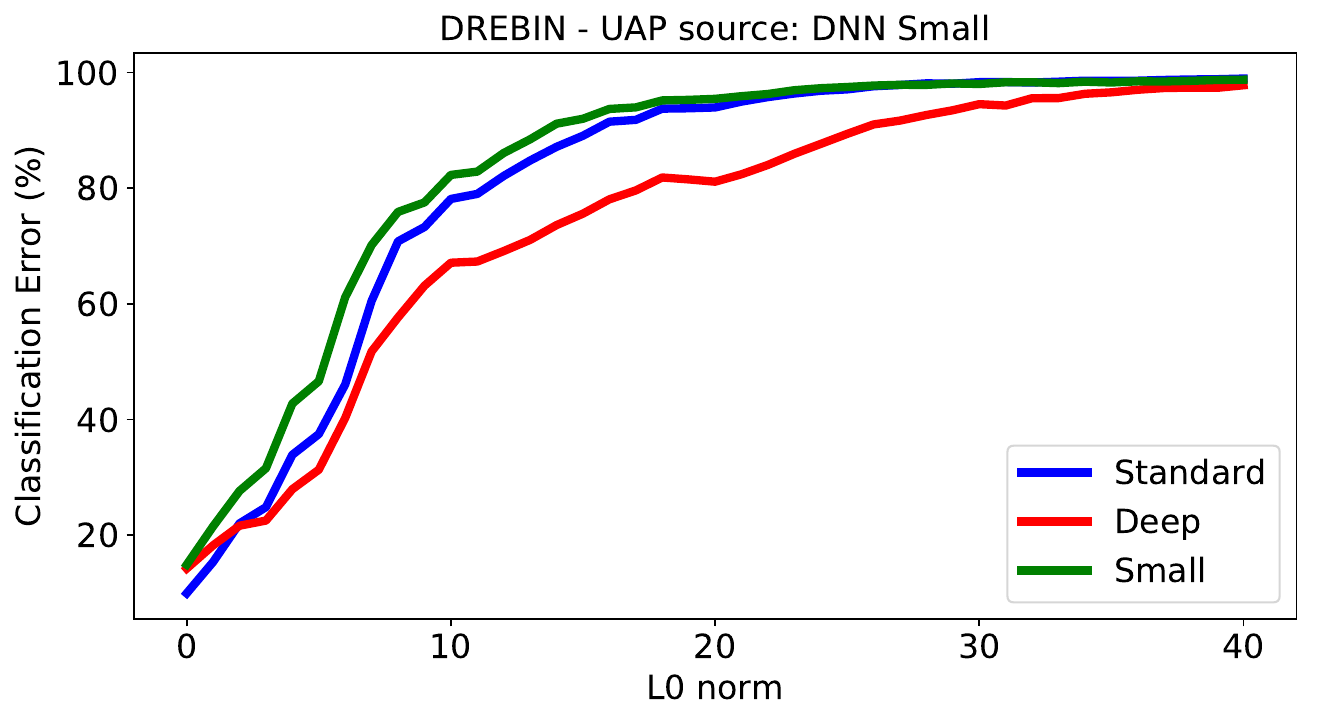} \\
    (b) Android \\
    \includegraphics[width= \columnwidth]{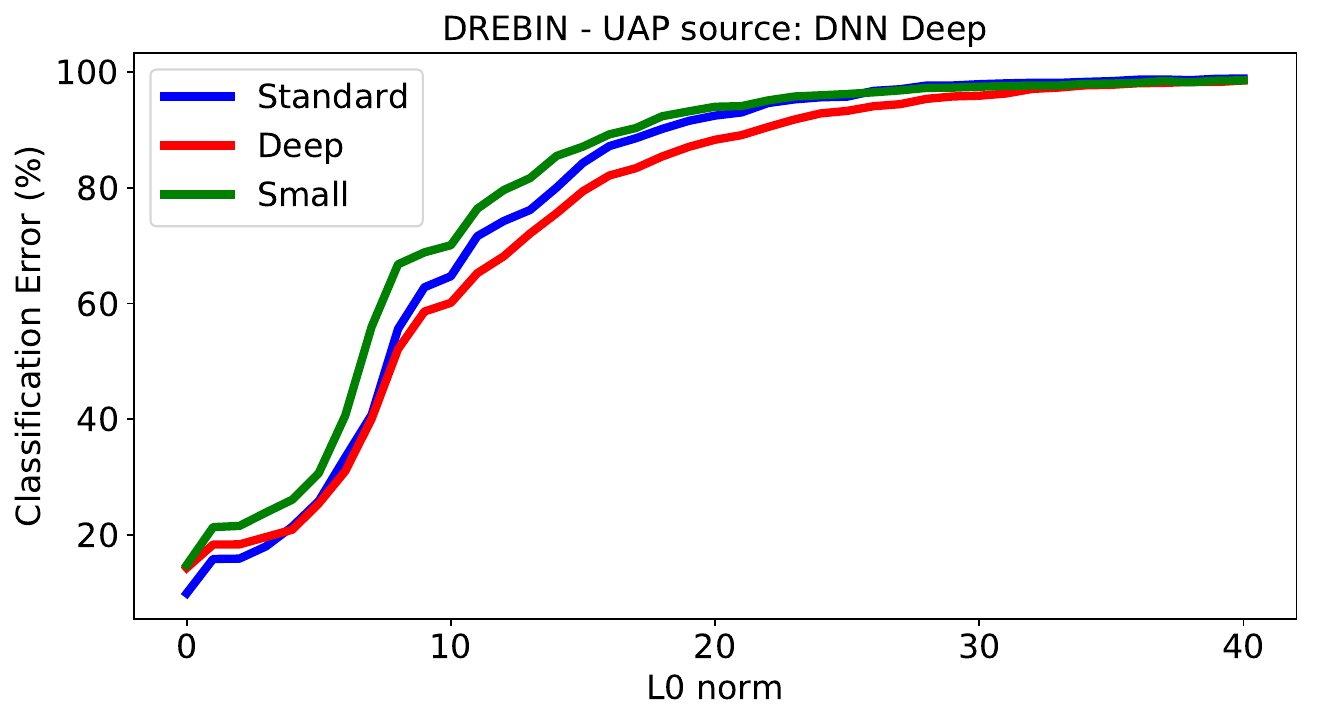} \\
    (c) Android \\
    \caption{Android malware (DREBIN): Transfer attacks using UAPs from (a) the standard DNN, (b) the small DNN, and (c) the deeper DNN.}
    \label{fig:transferabilityDREBINdnn}
\end{figure}

\begin{figure}
    \centering
    \includegraphics[width= \columnwidth]{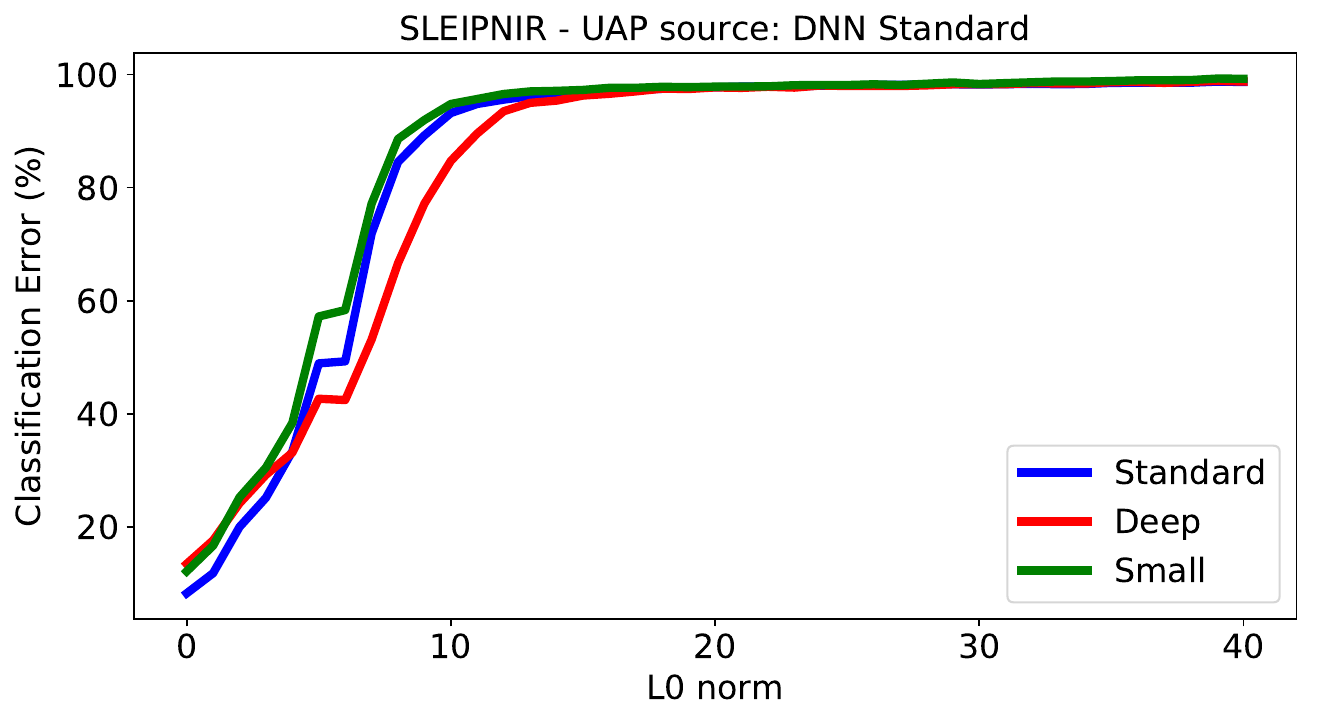} \\
    (a) Windows \\
    \includegraphics[width= \columnwidth]{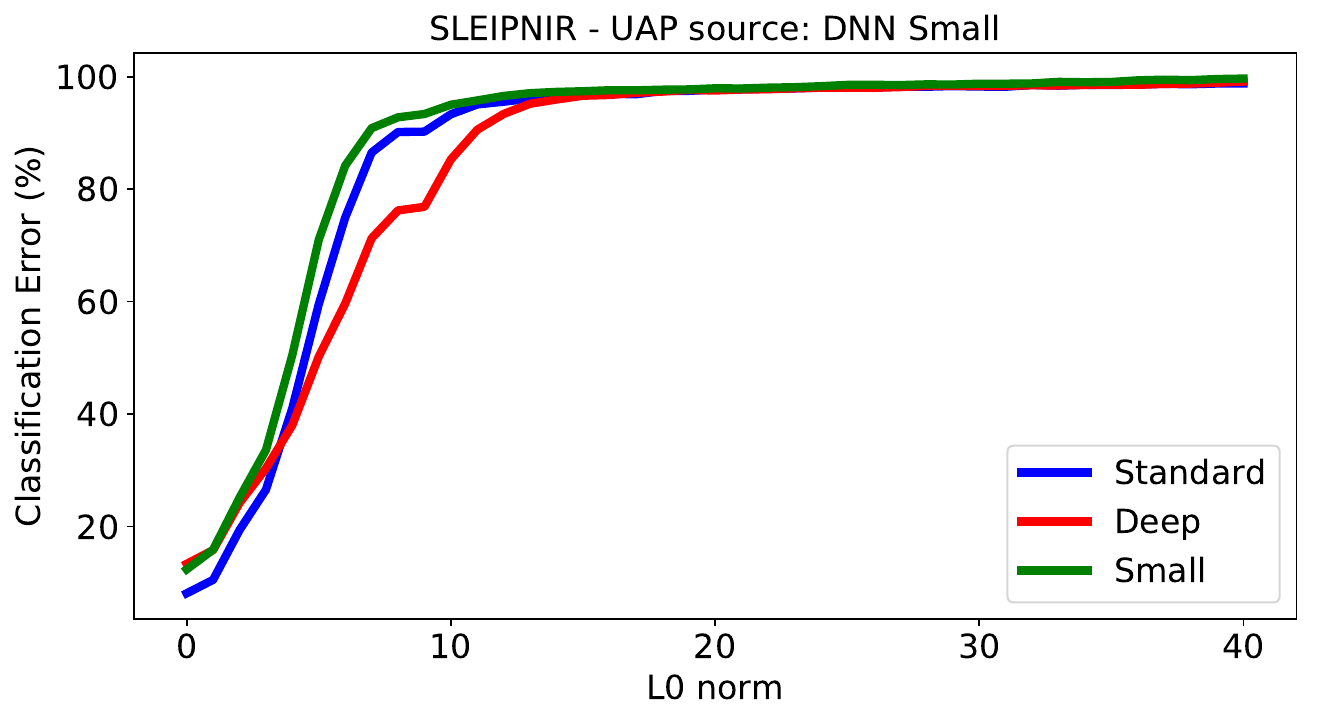} \\
    (b) Windows \\
    \includegraphics[width= \columnwidth]{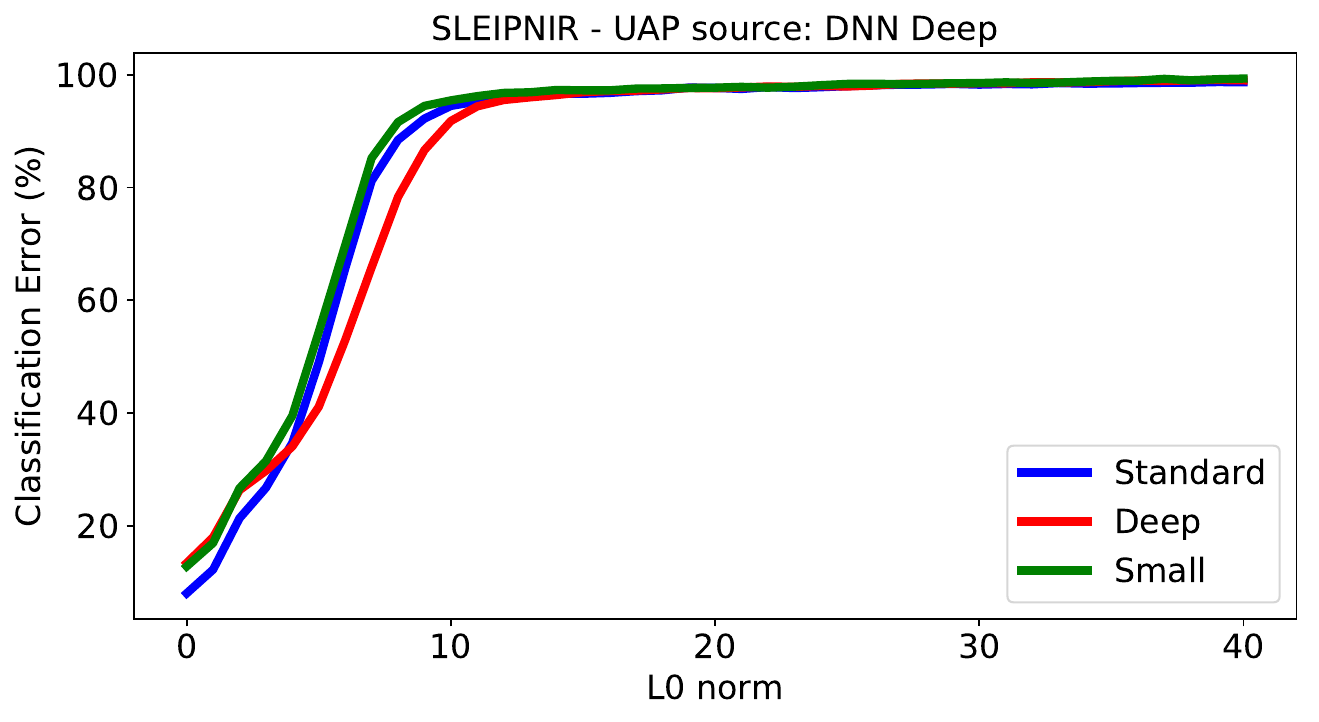} \\
    (c) Windows \\
    \caption{Windows malware (SLEIPNIR): Transfer attacks using UAPs from (a) the standard DNN, (b) the small DNN, and (c) the deeper DNN.}
    \label{fig:transferabilitySLEIPNIRdnn}
\end{figure}

As in \cref{section:transferability} we generate UAPs using the three models, and then perform transfer attacks. The results in \Cref{fig:transferabilityDREBINdnn,fig:transferabilitySLEIPNIRdnn} show that the deep model is slightly more robust than the other two (especially for DREBIN), but the improvements are marginal and for $L_0=40$ the UER for both the white-box and transfer attacks is almost 100\% in all cases and for all models. We also observe that the attack transferability is very good, especially for the standard and small model and the transfer attacks are almost as effective as the white-box ones. 

Our experimental results show that the model architecture does not have a significant impact on the robustness of the classifier to UAP attacks, and the high transferability of the attacks enables black-box attacks. 

\begin{figure}[t]
    \centering
    \includegraphics[width=\columnwidth]{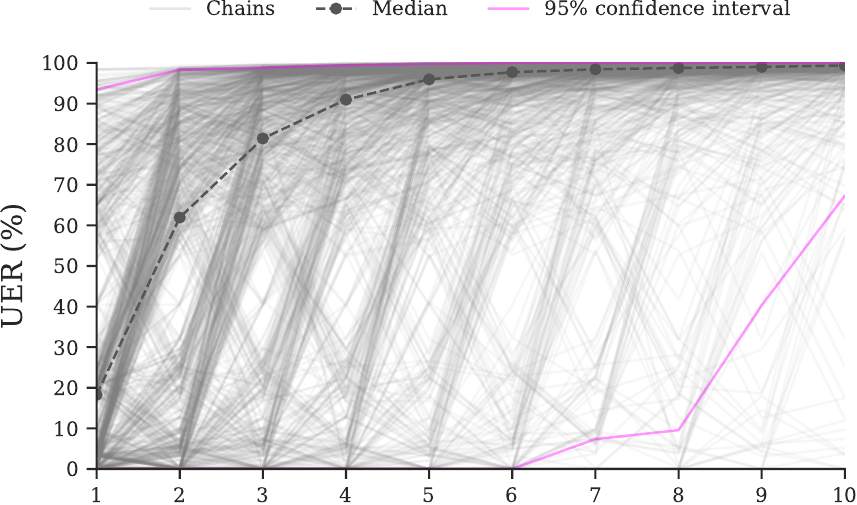}
    \caption{Limited Knowledge (LK) attack against a nonlinear \drebin Android malware classifier.}
    \label{fig:random-uer-test-nn}   
\end{figure}

\section{Additional Android Result}
\label{app:android-dnn-pk}

\Cref{fig:random-uer-test-nn} reports Limited Knowledge (LK) attack against a nonlinear \drebin Android malware classifier, implemented as a DNN with 2 hidden layers. Universal Evasion Rates (UER) are shown for 1,000 random transformation chains up to length 10. Relatively few transformations are required to achieve a high UER, highlighted by the median at each stage. Compared to the linear model~(\Cref{fig:random-uer-test-linear}), the nonlinear model seems even more susceptible to the LK attack, however the flexibility of the model enables us to devise a defense which is better able to adapt to the attacker's set of transformations.

\balance

\section{UAP Search with Genetic Programming}
\label{app:extra-gp-result}

As shown in Table~\ref{tab:uer-windows-gp}, we also evaluate genetic programming (GP) as a UAP search strategy. We observe that the UER is strongly affected compared to the earlier approach. Since GP gives more importance to chains, rather than focusing on individual transformations, distinct UAPs can be found, but GP tends to emphasize sequences that repeat transformations leading to a higher rates of corruption and otherwise reducing the plausibility of the binary. Similar behavior has also been reported using reinforcement learning for input-specific evasion attacks where the agent tends to overuse actions and generate repeated sequences of transformations~\cite{anderson2018learning,labaca2021aimed}. 

\begin{table*}[hbt]
\centering
\footnotesize
\caption{Adaptive attack in the Windows domain (UER at $|\seqT|$ of 4 and 10) using a genetic programming (GP) approach for UAP search.}
\scalebox{1.2}{ %
\begin{tabular}{clrrrrrr}
  \toprule 
& {\sc Model} & {\sc Search} & {\sc AUC-ROC} & {TPR} & {\sc $UER_1$} & {\sc $UER_4$} & {\sc $UER_{10}$} \\
  \midrule 
\multirow{3}{*}{} %
& \ember-LGBM $C = .90$ & GP & 0.999 & 0.921 & --\% & 0.6\% & 3.5\%  \\
& \ember-LGBM $C = .87$ & GP & 0.999 & 0.930 & --\% & 0.0\% & 0.1\%  \\
\bottomrule
\end{tabular}}
\label{tab:uer-windows-gp}
\end{table*}

\end{document}